\documentclass[final]{siamltex}
\usepackage{times}
\usepackage{graphics,epsfig,graphicx,color}
\usepackage{amsmath}
\usepackage{mathrsfs,amsfonts,amssymb}
 \usepackage{bm}
\usepackage{url}
\DeclareMathAlphabet{\itbf}{OML}{cmm}{b}{it}
\def\EE{\mathbb{E}}

\def\RR{\mathbb{R}}

\def\la{\lambda}

\def\bX{{\itbf X}}
\def\bx{{\itbf x}}
\def\bu{{\itbf u}}
\def\by{{\itbf y}}
\def\btheta{{\boldsymbol{\theta}}}

\def\hu{{\hat{\vec{\boldsymbol{\mathfrak{u}}}}}}

\newcommand{\dessous}[2]{
\renewcommand{\arraystretch}{0.5} 
\begin{array}[t]{c}
{#1} \\
\scriptstyle
{#2}
\displaystyle
\end{array}
\renewcommand{\arraystretch}{1.0}
}

\newtheorem{alg}[theorem]{Algorithm}
\begin{document}

\title{Laser beam imaging from the speckle pattern of the off-axis
  scattered intensity}

\author{
Liliana Borcea\footnotemark[1]
 and
Josselin Garnier\footnotemark[2]
}

\maketitle

\renewcommand{\thefootnote}{\fnsymbol{footnote}}

\footnotetext[1]{Department of Mathematics, University of Michigan,
  Ann Arbor, MI 48109. {\tt borcea@umich.edu}}
\footnotetext[2]{Centre de Math\'ematiques Appliqu\'ees, Ecole
  Polytechnique, 91128 Palaiseau Cedex, France.  {\tt
    josselin.garnier@polytechnique.edu}}

\renewcommand{\thefootnote}{\arabic{footnote}}

\begin{abstract}
  We study the inverse problem of localization (imaging) of a laser
  beam from measurements of the intensity of light scattered off-axis
  by a Poisson cloud of small particles. Starting from the 
  wave equation, we analyze the microscopic coherence of the scattered
  intensity and show that it is possible to determine the laser beam
  from the speckle pattern captured by a group of CCD cameras. Two
  groups of cameras are sufficient when the particles are either small
  or large with respect to the wavelength. For general particle sizes
  the accuracy of the laser localization with two groups of cameras is
  subject to knowing the scattering properties of the cloud. However,
  three or more groups of cameras allow accurate localization that is
  robust to uncertainty of the type, size, shape and concentration of
  the particles in the cloud. We introduce a novel laser beam
  localization algorithm and give some numerical illustrations in a
  regime relevant to the application of imaging high energy lasers in
  a maritime atmosphere.
\end{abstract}



\pagestyle{myheadings}
\thispagestyle{plain}
% \markboth{L. Borcea and J. Garnier}
% {Laser beam imaging}

\section{Introduction}
We study an inverse problem for the wave equation, motivated by the
application of detection and characterization of high energy laser
beams propagating in a maritime atmosphere. The data are gathered by
sensors that do not lie in the footprint of the beam, but at remote locations off its axis. These sensors measure
the intensity of incoherent light scattered by a cloud of particles
suspended in air (aerosols), with sizes ranging from a few nanometers
to a hundred micrometers \cite{kusmierczyk}.  Maritime environments
have a mixture of aerosols like sea salts, dust particles, water
droplets, etc., with concentration and composition of the cloud
depending on factors such as weather and location
\cite{korotkova,VanEijk}. The aerosols are typically modeled as
spherical particles, so that their interaction with the laser beam can be
described by the Mie scattering theory \cite{ishimaru,vandehulst}.
This seems to capture well experimental observations
\cite{cariou03,VanEijk,korotkova}.

We refer to \cite{cariou03,Roy,kusmierczyk} for studies of detection
of laser beams from off-axis measurements of the scattered
intensity. Pulsed laser beam localization is studied in
\cite{hanson11,WeiDong}, using a camera that can measure the intensity
resolved over both time and direction of arrival. We are
interested in continuous wave (time harmonic) laser beams, where
arrival times cannot be measured.  The localization of such lasers was
studied in \cite{hanson13} using intensity measurements at two cameras
placed in the focal plane of lenses which Fourier transform the
light wave in order to resolve the intensity over direction of
arrival. The setup requires knowledge of the focal length of the
lenses, which depends on the wavelength $\lambda$ of the laser.  The
localization becomes ambiguous when the two cameras and the axis of
the beam are in the same plane, as shown in \cite{hanson13}, where an
improvement based on the relative radiance of scattering at the
cameras is proposed. This approach may be susceptible to uncertainty
in the composition of the cloud of particles and of the mathematical
model of the scattered intensity.

In this paper we introduce an original method for laser beam
localization, using the microscopic coherence properties of the
intensity measured off-axis. It has the advantage of robustness to
uncertainty of the wavelength of the laser beam, the shape and size of
the particles, and the concentration and composition of the
cloud. However, it requires more measurements of the intensity, at two
or more groups of cameras, and these measurements must be spatially
resolved according to the speckle size, which is determined by the
dominant type of particles in the cloud. If most particles are small
with respect to the wavelength, the speckle size is of the order
$\la$, and the cameras may need to be equipped with microscopes for
proper spatial resolution. The speckle size increases for larger
particles, so conventional CCD cameras have sufficient resolution.

We derive from first principles, starting from the  wave equation, the
mathematical model of the intensity of the incoherent light scattered
off-axis by a cloud of particles encountered by the laser beam. The
locations of the particles are modeled by a Poisson point process,
which corresponds to having statistically independent numbers of
particles in non-overlapping domains. We begin with a Poisson cloud of
identical, spherical particles of radius $a$, and derive a simple
model of the scattered intensity using the single scattering (Born)
approximation and the Mie theory. This gives an explicit mathematical
expression of the incoherent intensity that shows the dependence of
the speckle pattern on the ratio $a/\la$. Then we explain how the
results generalize to mixtures of particles of different sizes and
shapes, and to multiple scattering regimes, as long as the waves
reaching the cameras do not travel longer than the transport mean free
path in the Poisson cloud. This is the characteristic length scale
over which the light forgets its initial direction due to multiple
scattering \cite{VanRossum}. At larger travel distances the angle of
arrival of the recorded intensity is not meaningful, and imaging
should be based on diffusion models.

In this paper we image at distances smaller than the transport mean
free path, and show how to extract information about the laser beam
from the  speckle pattern of the off-axis scattered
intensity. We introduce a novel imaging algorithm and analyze how many
measurements are needed for accurate beam localization that is robust
to uncertainty of the cloud of particles and therefore of the model
of the measurements.

The paper is organized as follows: We begin in section \ref{sect:form}
with the formulation of the problem and the scaling regime. Then we
give in section \ref{sect:stat} the statistics of the 
waves scattered off-axis, and describe in detail the covariance of the speckle
intensity. The imaging algorithm is introduced in section \ref{sect:imag}
and its performance is illustrated with some numerical simulations in
section \ref{sect:num}. We end with a summary in section \ref{sect:sum}.

\section{Formulation of the problem}
\label{sect:form}
We give here a simple model of the interaction of a laser beam with a
Poisson cloud of particles. We derive it in section
\ref{sect:Born}, using the single scattering approximation and the Mie
scattering theory, in  the scaling regime described in section
\ref{sect:scale}. 

\subsection{Model of the scattered waves}
\label{sect:Born}
Let us begin with the Helmholtz equation 
\begin{equation}
  \label{eq:form1}
\Delta u(\vec\bx) + (k+i k_{\rm d})^2 \big[1+ V(\vec\bx)\big]
u(\vec\bx) = 0 ,
\end{equation}
satisfied by a time harmonic wave $u(\vec\bx)e^{-i \omega t}$ at
frequency $\omega$ and location $\vec\bx \in \mathbb{R}^3$. The wave
propagates in a medium with constant wave speed $c$, containing small
particles modeled by the scattering potential $V(\vec\bx)$. The
coefficient $k$ in \eqref{eq:form1} is the wavenumber
\[
k = \frac{\omega}{c} = \frac{2 \pi}{\la},
\]
and $k_{\rm d}$ is a small damping parameter, satisfying $k \gg k_{\rm d}
> 0$, which models attenuation in the medium and extinction of the beam 
due to scattering by the cloud of particles \cite{VanEijk}.

The scattering potential $V(\vec\bx)$ is supported on the particles,
modeled as spheres $B(a_j, \vec\bx_j)$ of radius $a_j$ and center
$\vec\bx_j$, for $j \ge 1$,
\begin{equation}
V(\vec\bx) = \sum_j \sigma_j {\bf 1}_{B(a_j,\vec\bx_j)}(\vec\bx).
\label{eq:form2}
\end{equation}
Here ${\bf 1}_{B(a_j,\vec\bx_j)}$ is the indicator function of the support
of the $j$-th particle and $\sigma_j$ is its reflectivity, the change
in the index of refraction.  The locations $\{\vec\bx_j\}_{j \ge 1}$
of the particles are modeled as a Poisson point process with homogeneous
intensity $\rho$. This is the mean number of particles per unit
volume, and it can be written as
\begin{equation}
  \rho = 1/\ell^3,
  \label{eq:form.3}
\end{equation}
with $\ell$ interpreted as the mean distance between the particles.
We consider first identical particles with radius $a$, so that we can
study the effect of the ratio $a/\lambda$ on the speckle pattern
registered at the cameras. As explained later, the imaging method
applies to a mixture of particle sizes and shapes.

The wave field
\begin{equation}
u(\vec\bx) = u_{\rm b} (\vec\bx)+u_{\rm s}(\vec\bx)
\label{eq:form4}
\end{equation}
is the superposition of the incident field $u_{\rm b}(\vec\bx)$, which
models the laser beam, and the scattered field $u_{\rm s}(\vec\bx)$.  For
convenience in the calculations, we assume that the beam has a
Gaussian profile, with axis parametrized by $z$ and beam waist in the
plane $z=0$. The radius at the waist is denoted by $r_o$.  It is large
with respect to the wavelength, so we are in a paraxial regime with
the beam modeled by \cite[Chapter 5]{mandel}
\begin{equation}
u_{\rm b}(\vec\bx) = \frac{r_o^2}{R_z^2} \exp\Big( - \frac{|\bx|^2}{R_z^2}
+i kz - k_{\rm d} z\Big),\quad \quad R_z = r_o \Big(1 +\frac{2i z}{k
  r_o^2}\Big)^{1/2}.
\label{eq:form5}
\end{equation}
Here we introduced the system of coordinates $\vec\bx = (\bx,z)$, with
$z$ on the axis of the laser beam, and the two-dimensional vector $\bx$ in
the plane orthogonal to it\footnote{
We denote herein vectors in three dimensions by bold letters and arrows, 
and two dimensional vectors by bold letters. We also denote 
unit vectors by hats. If these  are three-dimensional, they are 
also denoted by arrows, as in $\hat{\vec\bu}$.}.

In the single scattering (Born) approximation, the scattered field is
modeled by the solution of the inhomogeneous Helmholtz equation
\begin{equation}
\label{eq:form6}
  \Delta u_{\rm s} + (k+ik_{\rm d})^2 u_{\rm s} = -(k+ik_{\rm d})^2
  V(\vec\bx) u_{\rm b}(\vec\bx),
\end{equation}
satisfying the Sommerfeld radiation condition away from the beam and
outside the  support of $V(\vec\bx)$. It is given explicitly by
\begin{equation}
  u_{\rm s}(\vec\bx) = (k+ik_{\rm d})^2 \int_{\RR^3} d \vec \by\, 
  G(\vec\bx,\vec\by) V(\vec\by) u_{\rm b}(\vec\by),
  \label{eq:form8}
\end{equation}
where 
\begin{equation}
  G(\vec\bx,\vec\by) = \frac{1}{4 \pi |\vec \bx-\vec \by|} \exp
  \big[ (i k - k_{\rm d}) |\vec \bx-\vec \by|\big]
  \label{eq:form7}
\end{equation}
is the Green's function.  Using the model \eqref{eq:form2} of
the scattering potential, we  rewrite \eqref{eq:form8} as a sum
over the particles
\begin{equation}
  u_{\rm s}(\vec \bx) \approx k^2 \sum_{j} {\mathfrak I}_{\rm
    Mie}\big(\alpha(\vec\bx, \vec\bx_j);ka,\sigma\big)
  G(\vec\bx,\vec\bx_j) u_{\rm b}(\vec\bx_j).
  \label{eq:form9}
\end{equation}
Here we neglected the small damping term $k_{\rm d}$ in the multiplicative factor
$(k+ik_{\rm d})^2$, and introduced the Mie scattering kernel
${\mathfrak I}_{\rm Mie}$ \cite[Chapter 9]{vandehulst}, which depends
on the ratio of the radius $a$ of the particles and the wavelength
(i.e., $ka$), the reflectivity $\sigma$ and the angle $\alpha(\vec
\bx,\vec \bx_j)$ from $\vec\bx$ to  $\vec \bx_j$.

For small (point-like) particles, with radius $a$ satisfying $ka \ll
1$, the scattering is approximately isotropic and we can approximate
the kernel ${\mathfrak I}_{\rm Mie}$ by a constant
\begin{equation}
  {\mathfrak I}_{\rm Mie}\big(\alpha(\vec\bx,
  \vec\bx_j);ka,\sigma\big) \approx \sigma \frac{4 \pi a^3}{3} =: \eta.
  \label{eq:form10}
\end{equation}
When\footnote{We use
  throughout the symbol $\sim$ to denote of the order of, the symbol
  $\gtrsim$ to denote larger or at least of the order of, and the
  symbol $\lesssim$ to denote smaller or at most of the order of.}
 $k a \gtrsim 1$ but $\sigma$ is small enough
so that $\sigma k a \ll 1$, the scattering kernel is
approximated by the Rayleigh-Gans formula \cite[Chapter 7]{vandehulst}
\begin{equation}
   {\mathfrak I}_{\rm RG}\big(\alpha(\vec\bx,
  \vec\bx_j);ka,\sigma\big) =  \eta
  \frac{3 \sqrt{2 \pi} J_{3/2}\big[2 k a \alpha(\vec\bx,
  \vec\bx_j)\big]}{2 \big[2 k a \alpha(\vec\bx,
  \vec\bx_j)\big]^{3/2}},
  \label{eq:form11}
\end{equation}
where $J_{3/2}(t) = \sqrt{2/\pi} (\sin(t) -t\cos(t))/t^{3/2}$ is the
Bessel function of the first kind and of order $3/2$. The expression
\eqref{eq:form11} reduces to \eqref{eq:form10} in the limit $k a \to
0$, and shows that scattering is peaked in the forward direction, at
angles $\alpha \sim 1/(ka)$, when $ka \gtrsim 1$. The forward scattering
is also predicted by the Mie scattering kernel ${\mathfrak I}_{\rm
  Mie}$, which should be used for larger $\sigma$. This has a complicated
expression given in \cite[Chapter 9]{vandehulst}.

\subsection{Scaling}
\label{sect:scale}
Our analysis of the statistics of the scattered field \eqref{eq:form9}
is carried out in a regime defined by the relations\begin{equation}
  \la \ll  %A \lesssim
   \ell \ll r_o \ll L_\bx \ll L_z, \qquad
   \lambda \ll d_A \ll r_o ,
  \label{eq:form12}
\end{equation}
between the important length scales in the problem: the wavelength
$\la$, the particle size $a$, the mean distance $\ell$ between the
particles, the radius $r_o$ of the laser beam, the diameter $d_A$ of the domain (aperture) $A$ of the
camera, the typical offset (cross-range) $L_\bx$ of the camera from
the axis of the beam and the typical distance (range) $L_z$ of the
camera along the axis of the beam, measured from the waist (the laser
source).  

The scaling relations \eqref{eq:form12} are motivated by the
application of high energy laser imaging in a marine atmosphere, where
the wavelength $\la$ is of the order of $1\mu{\rm m}$, and the
particle radius $a$ may be small or large with respect to $\la$. The
mean distance $\ell$ between the particles is of the order of
$1$mm. It is much larger than the wavelength, so multiple scattering
is not too strong and the Born approximation captures approximately
the microscopic coherence properties of the speckle pattern. The
radius of the beam $r_o$ is in the range of $0.1-1$m. The diameter $d_A$
of the camera is of the order of hundreds of wavelengths. It is at
cross-range $L_\bx$ of the order of $100$m and at range $L_z$ of the
order of $1$km.

In this scaling regime, the Rayleigh length $L_{\rm R}$, which is the
distance at which the beam doubles its radius due to diffraction,
satisfies
\begin{equation}
  L_{\rm R} = \frac{k r_o^2}{2} \gg L_z.
  \label{eq:form13p}
\end{equation}
Thus, we may neglect diffraction effects and approximate in \eqref{eq:form5}
\begin{equation}
  R_z \approx r_o~~ \mbox{for} ~ z = O(L_z).
  \label{eq:form14}
\end{equation}
Nevertheless, it is possible to extend the results to scalings where $L_{\rm R}\sim L_z$ 
and $R_z$ is a  smooth  $z$-dependent function, as defined in (\ref{eq:form5}).

The damping term $k_{\rm d}$, which models attenuation in the medium, is
used in our analysis to ensure the integrability of the terms in the
sum \eqref{eq:form9}. We assume henceforth that
\begin{equation}
  k_{\rm d} L_\bx \ll 1,
  \label{eq:form13}
\end{equation}
so we can neglect the attenuation over the cross-range offsets from
the laser axis to the cameras. This assumption simplifies the
expression of the correlation function of the intensity of the
scattered field, derived in the next section. The results extend to
$k_{\rm d} L_\bx \gtrsim 1$, but from the practical point of view the
intensity may be too weak to be detected by such remote cameras.

%% Assuming that the source of the laser is in the angle of view of the
%% camera, and the axis of the beam is not parallel to the planar camera
%% aperture, we limit in our calculations the support of the Poisson
%% cloud to the range interval
%% \begin{equation}
%%   z \in (0, Z_{\rm P}), \quad L_\bx \ll Z_{\rm P} - Z \ll Z \sim L_z.
%%   \label{eq:assumez}
%% \end{equation}
%% The camera does not record the intensity of light reflected beyond the
%% range $Z_{\rm P}$, because that light comes from behind. We also
%% suppose for convenience that
%% \begin{equation}
%%   1 \ll   \frac{Z_{\rm P}-Z}{L_\bx} \ll \frac{1}{k_{\rm d} L_\bx},
%%   \label{eq:assumezp}
%% \end{equation}
%% so we can neglect the attenuation effects over the range offset
%% $Z_{\rm P}-Z$.

\section{Statistics of the scattered waves}
\label{sect:stat}
We describe here the statistics of the scattered wave field
$u_{\rm s}(\vec\bx)$ modeled by equation \eqref{eq:form9}. We begin in
section \ref{sect:poiss} with a summary of basic results for Poisson
point processes. Then we derive in section \ref{sect:small} the
expression of the covariance function of the intensity
$|u_{\rm s}(\vec\bx)|^2$ measured at the camera, for the case of small
particles. The case of larger particles is analyzed in section
\ref{sect:large}, and the generalization to mixtures of particles is
in section \ref{sect:general}. We  also analyze in section \ref{sect:level}
the level sets of the covariance function near its peak, and show that they can 
be approximated by ellipsoids with axes that depend on the axis 
of the laser beam. This is used in the imaging algorithm described in 
section \ref{sect:imag}.

\subsection{Basic results on Poisson point processes}
\label{sect:poiss}
Recall from section \ref{sect:Born} that the locations
$\{\vec\bx_j\}_{j\ge 1}$ of the particles are modeled by a Poisson
cloud with homogeneous intensity $\rho$. Here we summarize from
\cite{kingsman} some basic results on Poisson processes, needed to
calculate the statistical moments of the scattered wave field.

By Campbell's theorem \cite[Section 3.2]{kingsman}, for any function
$f(\vec\bx)$ satisfying the condition $\min(|f|,1)\in L^1(\RR^3)$, the
characteristic function of the random variable $F=\sum_j f(\vec\bx_j)$
is given by
\begin{equation}
\EE [ e^{ i t F}] = \exp \Big[ \rho\int_{\RR^3} d\vec\bx \, \big( e^{it
    f(\vec\bx)} -1 \big) \Big],
\end{equation}
where $\EE[\cdot]$ denotes expectation with respect to the Poisson point process
distribution.  Moreover, if the function $f$ is in $L^1(\RR^3) \cap
L^2(\RR^3)$, then $F=\sum_j f(\vec\bx_j)$ is an integrable and
square-integrable random variable with
\begin{equation}
  \label{eq:expectation}
\EE[F] = \rho \int_{\RR^3} d\vec\bx\, f(\vec\bx), \quad\quad
\EE[F^2] = \rho \int_{\RR^3} d\vec\bx\, f^2(\vec\bx).
\end{equation}
The following lemma allows us to calculate the moments of the
scattered wave field:

\vspace{0.05in}
\begin{lemma}
Let $f_1,\ldots, f_4$ be functions in $L^1(\RR^3) \cap L^2(\RR^3)$,
that integrate to zero
\begin{equation}
  \label{eq:meanzero}
  \int_{\RR^3} d \vec\bx \, f_q(\vec\bx) = 0, \quad q = 1, \ldots, 4,
\end{equation}
and denote $ F_q=\sum_j f_q(\vec\bx_j) \,$ for $q = 1, \ldots, 4$.  We
have
\begin{equation}
\label{eq:camp2a}
\EE [ F_1 F_2 ] = \rho\int_{\RR^3} d\vec\bx \, f_1 (\vec \bx)
f_2(\vec \bx),
\end{equation}
and
\begin{align} \EE [ F_1 F_2 F_3 F_4 ]
=& \EE[ F_1 F_2 ] \EE[ F_3 F_4 ] + \EE[ F_1 F_3] \EE[ F_2 F_4 ] + \EE[
  F_1 F_4 ] \EE[ F_2 F_3 ] \nonumber \\ &+ \rho \int_{\RR^3} d\vec\bx\,
f_1(\vec\bx) f_2(\vec\bx) f_3(\vec\bx) f_4(\vec \bx).
\label{eq:camp2b}
\end{align}
\end{lemma}

\noindent
\begin{proof}
The proof follows from the identity
\begin{equation}
  \label{eq:pf1}\EE \Big[ \prod_{q=1}^n F_q\Big] =
  (-i)^n \frac{\partial^n}{\partial t_1\cdots \partial t_n} \EE \Big[
    \prod_{q=1}^n e^{i t_q \sum_j f_q(\vec\bx_j)}\Big] \Big|_{t_1, \ldots, t_n =0}
\end{equation}
and  \cite[Corollary 3.1]{kingsman}, which states that 
\begin{equation}
\EE \Big[ \prod_{q=1}^n e^{i t_q \sum_j f_q(\vec\bx_j)}\Big] =
\exp\Big[ \rho\int_{\RR^3}d\vec\bx\, \big( e^{i \sum_{q=1}^n t_q
    f_q(\vec\bx)}-1\big)\Big].
\label{eq:pf2}
\end{equation}
Equation \eqref{eq:camp2a} is obtained by substituting \eqref{eq:pf2} in
\eqref{eq:pf1}, setting $n = 2$, and using
\eqref{eq:meanzero}. Similarly, equation \eqref{eq:camp2b} follows by
substituting \eqref{eq:pf2} in \eqref{eq:pf1} and setting $n = 4$.
\end{proof}

When the functions $f_q$ are bounded and compactly supported, as is
the case in the model \eqref{eq:form9}, we note that the last term in
\eqref{eq:camp2b} is negligible with respect to the others if the
volume of support of the functions is large compared to $1/\rho =
\ell^3$. This condition holds in our scaling regime, and the
implication is that the fourth-order moments satisfy the Gaussian
summation rule, for zero-mean Gaussian processes. We use this
observation in the next sections and in appendix \ref{ap:Gaussian}, to
calculate the correlation of the intensity of the scattered field.

\subsection{Statistics of the scattered field for small scatterers}
\label{sect:small}
If the particles are small, with radius $a \ll \la$, the scattering
kernel in \eqref{eq:form9} is approximated by the constant $\eta$
defined in \eqref{eq:form10}.  The next proposition, proved in
appendix \ref{ap:propsmall}, gives the mathematical expression of the
mean and covariance function of the scattered field:

\vspace{0.05in}
\begin{proposition}
\label{prop:small}
In the scaling regime defined in section \ref{sect:scale}, the mean
scattered field at point $\vec \bx$ in the aperture of the camera is
approximately zero,
\begin{equation}
    \EE \big[ u_{\rm s}(\vec\bx)\big] \approx 0.
    \label{eq:prop1}
\end{equation}
Moreover, the covariance function of the scattered field evaluated at
points $\vec\bx_1=\vec\bX + \vec\bx/2$ and $\vec\bx_2=\vec\bX -
\vec\bx/2$ in the aperture of the camera is approximated by
\begin{equation}
\label{eq:cov1}
\EE \big[ u_{\rm s}(\vec\bx_1)\overline{u_{\rm s}}(\vec\bx_2) \big] \approx  \frac{
  \eta^2 k^4 \rho r_o^2 e^{-2k_{\rm d} Z}}{32 |\bX|} \Psi \Big(
k\hat{\bX} \cdot \bx,\frac{k r_o}{2 |\bX|} \hat{\bX}^\perp \cdot
\bx,kz \Big) ,
\end{equation}
where $\overline{u_{\rm s}}$ denotes the complex conjugate of $u_{\rm s}$ and 
\begin{equation}
  \label{eq:cov1p}
  \Psi( \chi,\xi,\zeta) = \frac{1}{\pi}\int_0^\pi \exp\Big[ i \big(
  \sin\alpha \chi+ \cos \alpha \zeta \big) \Big] \exp\Big( -
  \frac{\xi^2}{2}\sin^2 \alpha \Big) d\alpha .
\end{equation}
Here  we decomposed the vectors 
$\vec\bX = (\bX,{Z})$ and $\vec\bx = (\bx,z)$ in 
the range coordinates  $Z$ and $z$ along the axis of the laser beam, 
and  the two-dimensional vectors $\bX$ and $\bx$ in the cross-range plane, which is orthogonal to the beam. All coordinates are with respect 
to the origin that lies on the axis of the beam, at the waist. We also introduced the unit vector 
$\hat{\bX}= \bX/|\bX|$ and the unit vector $\hat{\bX}^\perp$, which is orthogonal to $\hat \bX$, and is defined 
by the rotation of
$\hat \bX$ by ninety degrees in the cross-range plane,
counterclockwise.
\end{proposition}

\begin{figure}[t]
\begin{center}
\begin{tabular}{c}
\hspace*{-0.25in}
\includegraphics[width=0.48\textwidth]{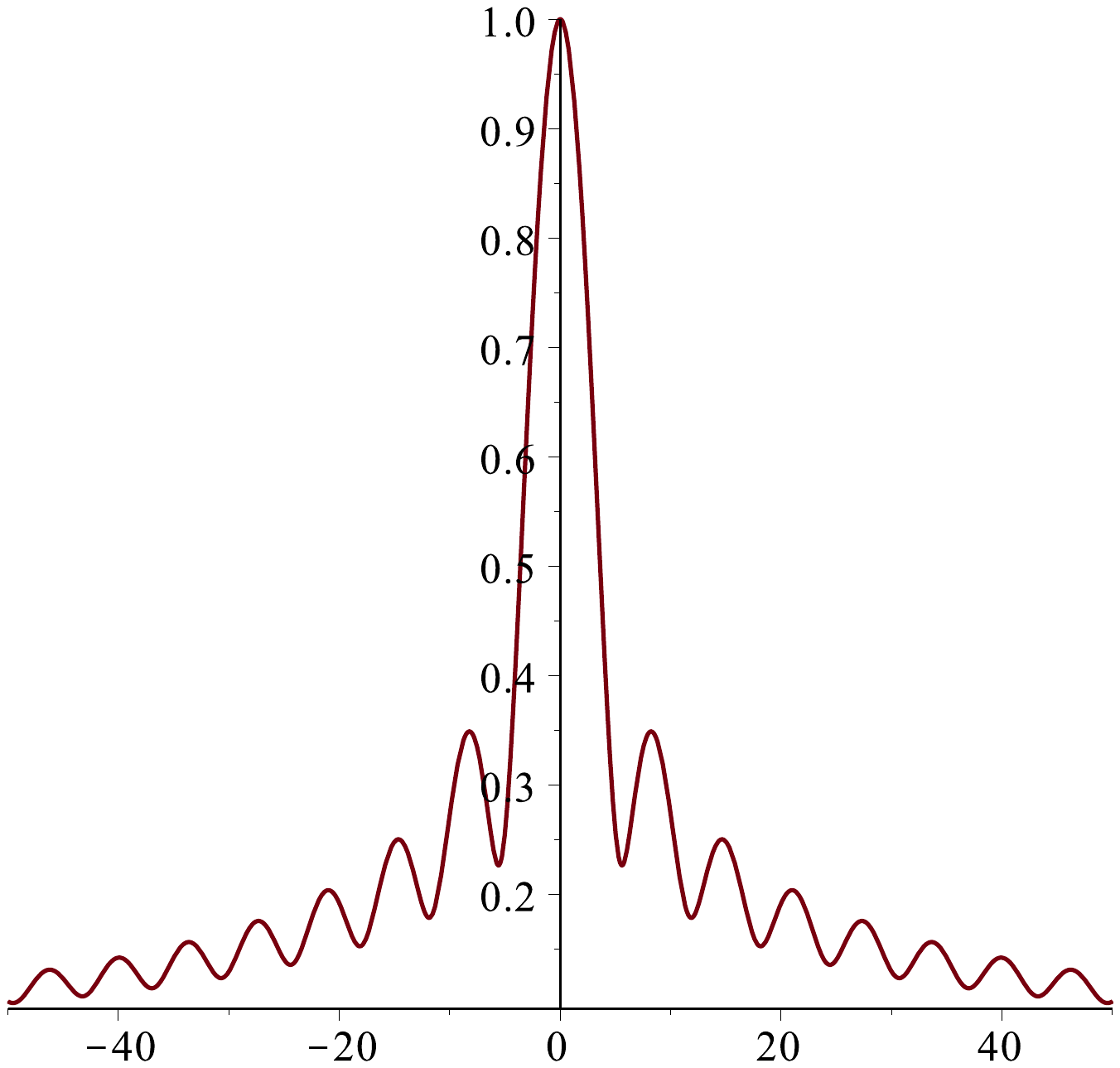}
\hspace*{-0.85in}
\includegraphics[width=0.48\textwidth]{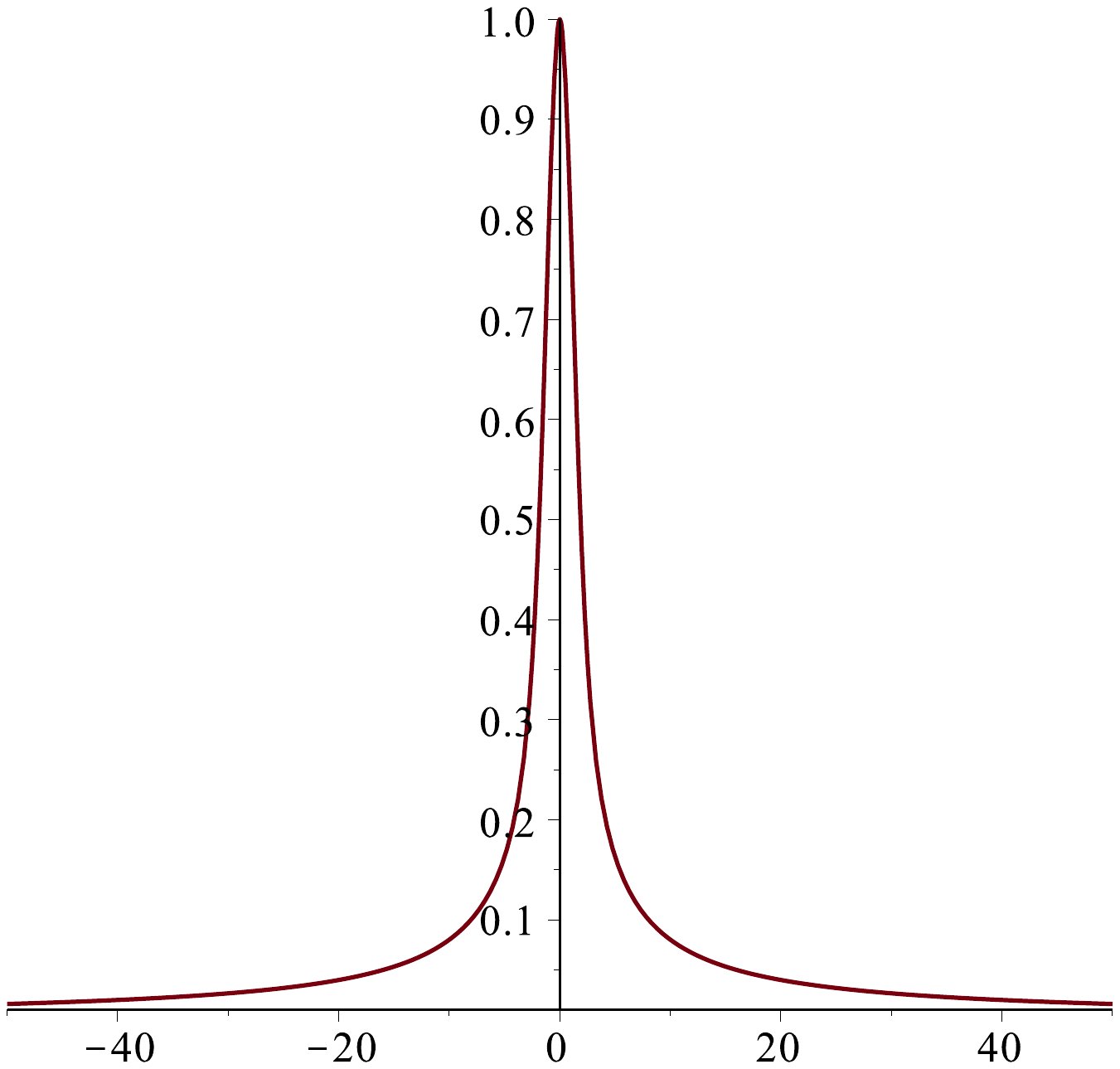}
\hspace*{-0.85in}
\includegraphics[width=0.48\textwidth]{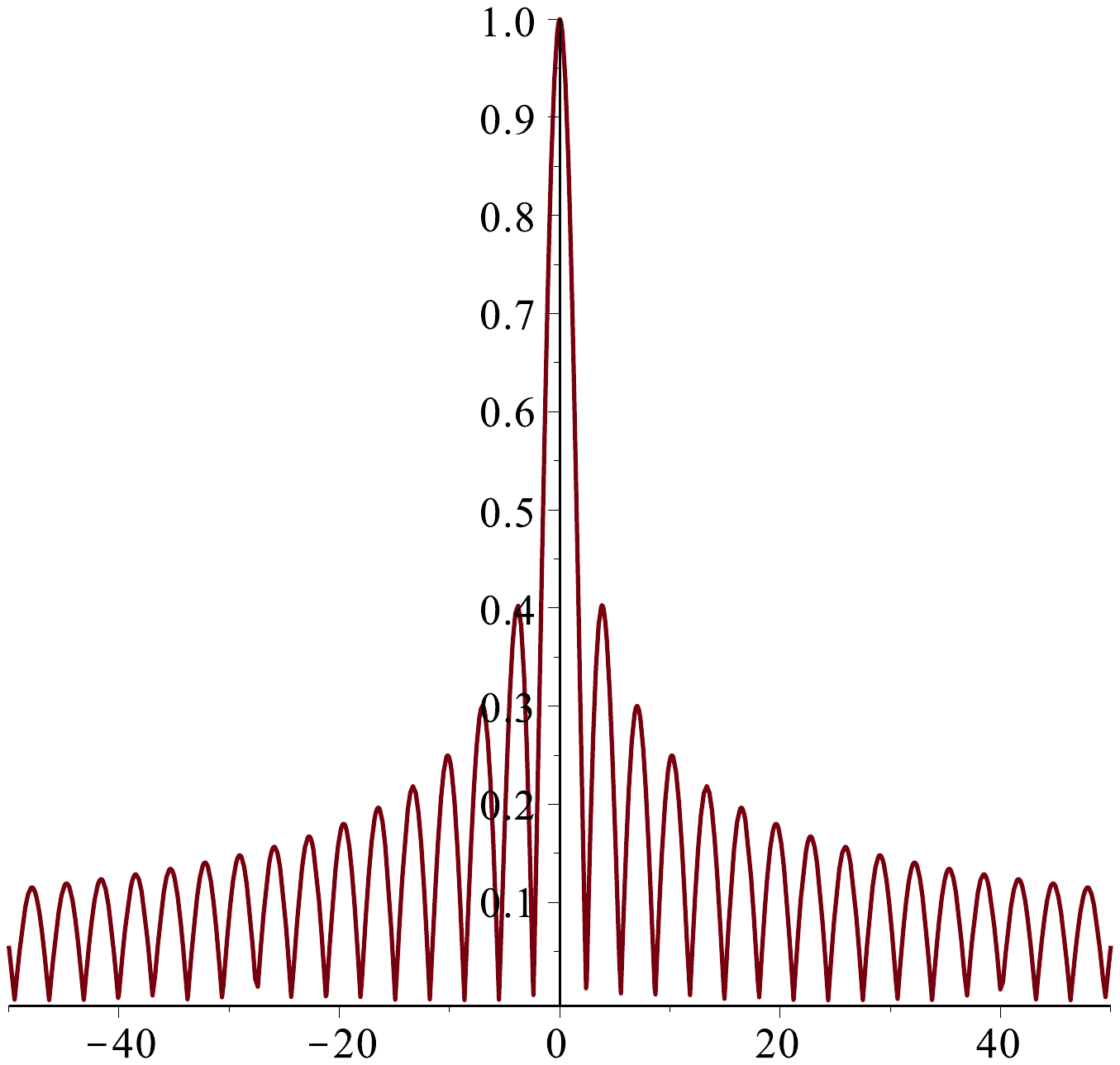}
\end{tabular}
\end{center}
\vspace{-1.2in}
\caption{From left to right we display functions $|J_0(t) + i H_0(t)|$,
  $I_0(t^2/4)e^{-t^2/4}$ and $|J_0(t)|\,$, for $|t| \le 50$.}
\label{fig:resol}
\end{figure}

\vspace{0.05in} There are two observations drawn from this
proposition: The first is that the scattered field at the camera is
incoherent, because its mean \eqref{eq:prop1}  is very
small with respect to its standard deviation that is approximately
equal to the square root of the mean intensity
\begin{equation}
\EE \big[ |u_{\rm s}(\vec\bX)|^2] 
=  \frac{ \eta^2 k^4 \rho r_o^2  e^{-2k_{\rm d} Z}}{32 |\bX|}.
\label{eq:meanInten}
\end{equation}
The second observation is that the second moment \eqref{eq:cov1},
which approximates the covariance of $u_{\rm s}$, has an anisotropic decay
that depends on the orientation of the axis of the laser beam.  To
estimate the decay of \eqref{eq:cov1} away from the peak, which occurs
at $\vec \bx_1 = \vec \bx_2$, we consider offsets $\vec \bx = \vec
\bx_1 - \vec \bx_2$ aligned with either one of the unit vectors $(\hat
\bX,0)$ and $(\hat \bX^\perp,0)$ in the cross-range plane, or with the range
axis. We have three cases:

\vspace{0.03in}
\begin{enumerate}
  \itemsep 0.03in
\item If $\vec \bx = |\bx|(\hat \bX,0)$, the covariance decays like
\begin{equation*}
\Psi( k|\bx|,0,0) = J_0( k| \bx|) + i H_0( k|\bx|),
\end{equation*}
where $J_0$ is the Bessel function of the first kind and of order
zero, and $H_0$ is the Struve function of order zero \cite[Chapter
  12]{abra}.
\item If $\vec \bx = |\bx|(\hat \bX^\perp,0)$, the covariance decays
  like
\begin{equation*}
  \Psi\Big(0,\frac{k r_o |\bx|}{2|\bX|},0\Big) = I_0\left[\frac{1}{4}
    \Big(\frac{k r_o |\bx|}{2|\bX|}\Big)^2\right] \exp
  \left[-\frac{1}{4} \Big(\frac{k r_o |\bx|}{2|\bX|}\Big)^2\right]
\end{equation*}
where $I_0$ is the modified Bessel function of the first kind and of
order zero \cite[Chapter 12]{abra}.
\item If $\vec \bx = z (0,0,1)$, the covariance decays like
\[
  \Psi(0,0,k z) = J_0(k z).
\]
\end{enumerate}
We plot in Figure \ref{fig:resol} the functions $|J_0(t) + i H_0(t)|$,
$I_0(t^2/4)e^{-t^2/4}$ and $|J_0(t)|\,$ and note that they are large
when the argument $t$ is order one.  Thus, we estimate that the
covariance decays on a scale comparable to the wavelength along the
cross-range direction $(\hat \bX,0)$ and the range direction
$(0,0,1)$. The decay in range is faster because as shown in Figure
\ref{fig:resol}, the support of the main peak of $|J_0(t)|$ is smaller
than that of $|J_0(t) + i H_0(t)|$ by a factor of approximately $2
\pi$. The decay of the covariance in the other cross-range direction
$(\hat \bX^\perp,0)$ is much slower, on the scale $|\bX| \la/r_o \gg
\la$.

The covariance function \eqref{eq:cov1} cannot be calculated directly,
because the camera does not measure the wave field $u_{\rm s}(\vec\bx)$, but
its intensity $|u_{\rm s}(\vec\bx)|^2$. The following proposition, proved in
appendix \ref{ap:Gaussian}, shows that the covariance of the measured
intensity is approximately the square of the modulus of
\eqref{eq:cov1}.
\vspace{0.05in}
\begin{proposition}
  \label{prop:Gauss}
  In the scaling regime described in section \ref{sect:scale}, and for
  two points $\vec\bx_1=\vec\bX + \vec\bx/2$ and $\vec\bx_2=\vec\bX -
  \vec\bx/2$ in the aperture of the camera, we have
\begin{equation}
\label{eq:approxGauss}
  {\rm Cov} \big( |u_{\rm s}(\vec\bx_1)|^2 , |{u_{\rm s}}(\vec\bx_2) |^2\big)
  \approx \big|\EE \big[ u_{\rm s}(\vec\bx_1)\overline{u_{\rm s}}(\vec\bx_2) \big]
  \big|^2.
\end{equation}
\end{proposition}

The covariance of the intensity can be estimated from the speckle
pattern captured by the camera, as explained in section \ref{sect:getcorrel}, 
and Propositions \ref{prop:small} and 
\ref{prop:Gauss}  give that we can use it to extract information about
the laser beam. The size of the speckles is related to the scales of
decay of the covariance, called correlation lengths.  The discussion
after Proposition \ref{prop:small}  shows that the
correlation lengths $l_{_\bX}$ and $l_{_Z}$ in the directions of the unit vectors
$(\hat \bX,0)$ and $(0,0,1)$, which span the plane containing $\vec{\bX}$ and the axis of the 
laser beam, are 
\begin{equation}
  l_{_{\bX}} \sim \la,  \quad l_{_Z} \sim \la, ~ ~ \mbox{such that} ~ ~
  l_{_{\bX}} > l_{_Z}.
\label{eq:correlsmall}
\end{equation}
The correlation length in the direction $(\hat \bX^\perp, 0)$
orthogonal to this plane is much larger
\begin{equation}
l_{_{\bX}}^\perp \sim
\frac{\la |\bX|}{r_o} \gg \la.
\label{eq:correlsmallp}
\end{equation}
The distance $|\bX|$ from the camera\footnote{We show later, in Lemma \ref{lem:replaceX}, that because the diameter $d_A$ of the aperture of the camera
is small,  the midpoint $\vec\bX$ may be replaced by the center of the camera.}  to the axis of the laser enters
the expression  \eqref{eq:cov1} of the covariance in the amplitude
factor and the correlation length $l_{_{\bX}}^\perp$. In practice, the
estimation should not be based on the amplitude, which depends on the
model and also on unknown parameters like $\eta$, $\rho$ and $k_{\rm
  d} Z$. Moreover, $l_{_{\bX}}^\perp$ is difficult to estimate from
the speckle pattern captured at a small camera with diameter $d_A \lesssim l_{_{\bX}}^\perp $.  
Thus, we do not estimate 
$|\bX|$ directly from the covariance function \eqref{eq:approxGauss}.

\subsection{Statistics of the scattered field for large scatterers}
\label{sect:large}
The results stated in the previous section extend readily to the case
of larger particles of spherical shape. The only difference in the
calculations, which are as in appendices \ref{ap:propsmall} and
\ref{ap:Gaussian}, is that the scattering kernel is no longer the
constant $\eta$, but a function rewritten here in the normalized form
\begin{equation}
  {\mathfrak I}_{\rm Mie}(\alpha; ka, \sigma) = \eta {\mathcal I}_{\rm
    Mie}(\alpha; ka, \sigma).
\end{equation}

\vspace{0.05in}
\begin{proposition}
\label{prop:large}
In the scaling regime defined in section \ref{sect:scale}, the mean
scattered field at the camera is approximately zero.  Moreover, the
covariance of the intensity at points $\vec\bx_1=\vec\bX + \vec\bx/2$
and $\vec\bx_2=\vec\bX - \vec\bx/2$ in the aperture of the camera is
given by the square of the modulus of the covariance of the scattered
field, as in \eqref{eq:approxGauss}. This covariance has the
mathematical expression
\begin{align}
  \label{eq:cov2a}
\EE \big[ u_{\rm s}(\vec\bx_1)\overline{u_{\rm s}}(\vec\bx_2) \big] &\approx
\frac{ \eta^2 k^4 \rho r_o^2e^{-2 k_{\rm d} Z}}{32 \pi |\bX|}
\Psi_{\rm Mie}\Big( k \hat{\bX} \cdot \bx ,\frac{k r_o}{2 |\bX|}
\hat{\bX}^\perp \cdot \bx, k z ; ka,\sigma \Big) ,
\end{align}
where 
\begin{align}
\Psi_{\rm Mie}( \chi ,\xi,\zeta; ka,\sigma)&=
\int_{0}^{\pi} \hspace{-0.05in} d \alpha \, \big|{\cal I}_{\rm
  Mie}\big( \alpha ; ka,\sigma \big)\big|^2 \exp\Big( i \big( \sin
\alpha \chi + \cos \alpha \zeta \big) - \frac{ \sin^2 \alpha}{2} \xi^2
\Big).
    \label{eq:cov2b}
\end{align}
\end{proposition}

The difference between the covariance functions \eqref{eq:cov1}
and \eqref{eq:cov2a} is the support of the kernel in the scattering
angle $\alpha$. While in the case of small particles the kernel is
constant, so that all angles $\alpha \in (0,\pi)$ contribute to the
integration in \eqref{eq:cov1p}, for larger particles only
smaller angles $\alpha$ contribute in \eqref{eq:cov2b} i.e.,
scattering is in the forward direction.

To illustrate the effect of forward scattering on the covariance
\eqref{eq:cov2a}, suppose that the particles are large such that $ka
\gg1$. Then, the angular opening $\Theta$ of the forward scattering
cone is small and we can simplify equations
(\ref{eq:cov2a}-\ref{eq:cov2b}) by changing the variable of
integration $\alpha \to \alpha \Theta$ and using the small argument
expansions of the exponent. We obtain
\begin{align}
\Big|\EE \big[ u_{\rm s}(\vec\bx_1)\overline{u_{\rm s}}(\vec\bx_2) \big]\Big|
&\approx \frac{ \eta^2 k^4 \rho r_o^2 e^{-2 k_{\rm d} Z}\Theta}{32 \pi
  |\bX| } \Big|\Psi_{\Theta}\Big( k \Theta \hat{\bX} \cdot \bx
,\frac{k r_o \Theta^2 }{2|\bX|} \hat{\bX}^\perp \cdot \bx, k \Theta^2
z \Big)\Big| ,
\label{eq:covarLarge}
\end{align}
where the function 
\begin{align}
\Psi_{\Theta}( \chi ,\xi,\zeta )&= \int_{0}^\pi d \alpha \, \big|{\cal
  I}_{\rm Mie} \big( \alpha \Theta ; ka,\sigma \big)\big|^2 \exp\Big(
i \big( \alpha \chi- \frac{\alpha^2}{2} \zeta \big)
-\frac{\alpha^2}{2}\xi^2 \Big)
\label{eq:psiTh}
\end{align}
peaks at the origin and has support of order one in all arguments.  To
be more explicit, consider the Rayleigh-Gans regime, where $\sigma$ is
so small that $\sigma k a\ll 1$. Then, the kernel in \eqref{eq:psiTh}
simplifies to 
\[
{\cal
  I}_{\rm Mie}\big( \alpha \Theta ; ka,\sigma \big) \approx
\frac{3 \sqrt{2 \pi}J_{3/2}(2 k a\alpha)}{2 (2 k a \alpha)^{3/2}},
\]
and the angular opening of the cone is $ \Theta \sim {1}/{(k a)}. $
With this estimate we conclude from \eqref{eq:covarLarge} that the
correlation lengths are
\begin{equation}
l_{_{\bX}} \sim \frac{1}{k \Theta} \sim a, \qquad l_{_{\bX}}^\perp \sim
\frac{|\bX|}{k r_o \Theta^2} \sim \frac{k a^2 |\bX|}{r_o}, \qquad
l_{_Z} \sim \frac{1}{k \Theta^2} \sim k a^2.
\label{eq:correlLarge}
\end{equation}
Like in the case of small particles, the largest correlation length is
$l_{_{\bX}}^\perp$. All correlation lengths are much larger than those
defined in \eqref{eq:correlsmall}--\eqref{eq:correlsmallp}, by the
factor $ka \gg 1$ in the  direction $(\hat \bX,0)$ and the
even larger factor $(ka)^2$ in the other directions. Furthermore, the
decay in the plane defined by the axis of the laser and the vector
$\vec \bX$ is more anisotropic, with $l_{_Z} \gg l_{_\bX}$.

\subsection{The level sets of the correlation function}
\label{sect:level}
Propositions \ref{prop:small}--\ref{prop:large} describe the dependence of the 
covariance function of the intensity on the unknown axis of the laser beam.
The amplitude of the covariance function is model-dependent, so we do not wish 
to base the imaging on it. We use instead the level 
sets of the covariance function near its peak, which have a generic dependence
on the axis of the laser beam, as we now explain. 

Let us define the correlation function of the intensity at points $\vec \bX \pm \vec \bx/2$, 
\[
{\rm Corr} \big( |u_{\rm s}(\vec \bX + \vec \bx/2)|^2,|{u_{\rm s}}(\vec \bX-\vec\bx/2)|^2\big) = 
\frac{{\rm Cov} \big( |u_{\rm s}(\vec \bX + \vec \bx/2)|^2,|{u_{\rm s}}(\vec\bX - \vec \bx/2)|^2\big)}{\EE \big[ |u_s(\vec{\bX} + \vec \bx/2)|^2\big] 
\EE \big[ |u_s(\vec{\bX} - \vec \bx/2)|^2\big]},
\]
and use the same decomposition $\vec \bX = (\bX,Z)$ and $\vec \bx = (\bx,z)$ of the mid-point and 
offset vectors as in Proposition  \ref{prop:small}.  The correlation function 
attains its maximum value $1$ when the two points coincide, and we  study its level 
sets for small offset vectors $\vec \bx$, decomposed as 
\begin{equation}
\vec \bx = x (\hat \bX, 0) + x^\perp (\hat \bX^\perp, 0) + z (0,0,1),
\label{eq:L1}
\end{equation}
in the orthonormal basis $\{(\hat \bX, 0), (\hat \bX^\perp, 0), (0,0,1)\}$, with two-dimensional unit vectors
$\hat \bX $ and $\hat \bX^\perp$ defined in Proposition \ref{prop:small}. If we 
scale the components of $\vec \bx$ by the characteristic correlation lengths,
\begin{equation}
x = \frac{1}{k} \chi, \quad x^\perp = \frac{2|\bX|}{k r_o} \xi,
\quad z = \frac{1}{k} \zeta,
\label{eq:Ldefcomp}
\end{equation}
we conclude from  Propositions \ref{prop:small}--\ref{prop:large} that 
\begin{align}
{\rm Corr} \big( |u_{\rm s}(\vec\bX + \vec \bx/2)|^2,|{u_{\rm s}}(\vec\bX -
\vec \bx/2)|^2\big)\approx & \Big| \int_{0}^{\pi} \hspace{-0.05in} d
\alpha \, \mathfrak{S}(\alpha) \exp\Big[ i \big( \sin \alpha \chi +
\cos \alpha \zeta \big)\Big]\nonumber \\ & \times \exp\Big(-
  \frac{\sin^2\alpha}{2} \xi^2 \Big)\Big|^2,
\label{eq:LCorrelMod}
\end{align}
where $\mathfrak{S}(\alpha)$ denotes the scattering kernel, normalized
so that $\int_0^\pi d \alpha \, \mathfrak{S}(\alpha) = 1.$ This kernel
is non-negative and 
proportional to $|\mathcal{I}_{\rm Mie}|^2$.

Consider a level set ${S}_{\mathfrak{L}}$ of the correlation
function at value $1 - {\mathfrak L}$, for $0 < {\mathfrak L} \ll
1$. We can approximate it by expanding \eqref{eq:LCorrelMod} about
$(\chi,\xi,\zeta) = (0,0,0)$, and obtain
\begin{align}
1- \mathfrak{L} &= {\rm Corr} \big( |u_{\rm s}(\vec\bX +\vec
\bx/2)|^2,|{u_{\rm s}}(\vec\bX-\vec \bx/2)|^2\big) \approx
1- (\chi, \xi, \zeta) {\bf H} \begin{pmatrix} \chi
  \\ \xi \\ \zeta \end{pmatrix},
\label{eq:LapproxEll}
\end{align}
for $\vec \bx \in S_{\mathfrak L}$ decomposed as in
\eqref{eq:L1}--\eqref{eq:Ldefcomp}. Here $-2 {\bf H} \in
\RR^{3 \times 3}$ is the Hessian of the correlation function at its maximum, 
with entries defined by
\begin{align*}
H_{11} &= \int_{0}^{\pi}\hspace{-0.05in} d \alpha \,
\mathfrak{S}(\alpha) \sin^2\alpha- \Big(
\int_{0}^{\pi} \hspace{-0.05in} d \alpha \,\mathfrak{S}(\alpha)
\sin\alpha \Big)^2, \\ H_{33}&= \int_{0}^{\pi}\hspace{-0.05in} d
\alpha \,\mathfrak{S}(\alpha) \cos^2\alpha - \Big(
\int_{0}^{\pi} \hspace{-0.05in} d \alpha \,\mathfrak{S}(\alpha)
\cos\alpha \Big)^2, \\ H_{13}&= \int_{0}^{\pi}\hspace{-0.05in} d
\alpha \,\mathfrak{S}(\alpha) \cos\alpha \sin \alpha - \Big(
\int_{0}^{\pi} \hspace{-0.05in} d \alpha \,\mathfrak{S}(\alpha)
\cos\alpha \Big)\Big( \int_{0}^{\pi} \hspace{-0.05in} d \alpha
\,\mathfrak{S}(\alpha) \sin\alpha \Big) ,\\ H_{22} &=
\int_{0}^{\pi} \hspace{-0.05in} d \alpha \,\mathfrak{S}(\alpha)
\sin^2\alpha, \\
H_{12} &= H_{23} = 0.
\end{align*}
Note that since the correlation function decays away from the peak at
$(\chi,\xi,\zeta) = (0,0,0)$, the matrix ${\bf H}$ is positive
definite.

Treating the approximation in \eqref{eq:LapproxEll} as an
equality, and recalling the scaling in \eqref{eq:L1}, we obtain
that the level set $S_{\mathfrak L}$ is the ellipsoid
\begin{equation}
 x^2 \left(\frac{k^2H_{11}}{\mathfrak{L}} \right) + z^2 \left(\frac{k^2 H_{33}}{\mathfrak{L}}\right)+2 x z
  \left(\frac{k^2 H_{13}}{\mathfrak{L}} \right) + (x^\perp)^2 \left[\Big(\frac{k r_o}{|\bX|}\Big)^2
    \frac{H_{22}}{{\mathfrak L}}\right] = 1,
\label{eq:Lellipse}
\end{equation}
with one principal axis along the unit vector $(\hat \bX^\perp,0)$
and the other principal axes in the plane containing the midpoint $\vec \bX$ and the axis of the laser beam, spanned 
by $(\hat\bX,0)$ and $(0,0,1)$. We distinguish three cases:

\vspace{0.03in}
\begin{enumerate}
  \itemsep 0.03in
\item When the particles are small with respect to the wavelength, so
  that scattering is isotropic (i.e, $\mathfrak{S} \equiv 1/\pi$), the
  ellipsoid is given explicitly by
  \[
\frac{x^2}{\big[\sqrt{{2}\mathfrak{L}/(1-8/\pi^2)}/{k}\big]^2} +
\frac{z^2}{(\sqrt{2\mathfrak{L}}/{k})^2} +
\frac{(x^\perp)^2}{\big[\sqrt{2\mathfrak{L}} |\bX|/(k r_o)\big]^2} =1,
\]
and its principal axes are along the basis vectors $\{(\hat \bX, 0), (\hat \bX^\perp, 0), (0,0,1)\}$. The largest axis is along
$(\hat \bX^\perp,0)$, and the smallest axis is along
$(0,0,1)$.
\item When the particles are large with respect to the wavelength, so
  that scattering is peaked forward, $\mathfrak{S}(\alpha)$ is
  supported in a cone of small opening angle $\Theta$. For example,
  $\Theta \sim 1/(ka) \ll 1$ in the Rayleigh-Gans regime, and the
  coefficients in \eqref{eq:ellipse} are estimated as
  \[
  H_{11} \sim \Theta^2, \quad H_{33} \sim \Theta^4, \quad H_{13} \sim
  \Theta^3, \quad H_{22} \sim \Theta^2.
  \]
  The ellipsoid \eqref{eq:Lellipse} has the largest principal axis
  along $(\hat \bX^\perp,0)$, and its length is larger than in the
  case of small particles, by a factor of order $1/\Theta$. The other
  axes  are no longer aligned with the basis vectors 
  $(\hat\bX,0)$ and $(0,0,1)$, but are rotated by a small
  angle of order $\Theta$. Their lengths are also much larger than in
  the case of the small particles. Moreover, the smallest principal
  axis is almost along $(\hat\bX,0)$.
\item For particles of intermediate size $a/\la \sim 1$, the
  coefficients $H_{11}, H_{33}, H_{13}, H_{22}$ are of the same
  order. Again, we conclude that the ellipsoid \eqref{eq:Lellipse} has
  the longest principal axis along $(\hat \bX^\perp,0)$. The other
  axes are in the plane containing $\vec{\bX}$ and the
  axis of the laser beam, but they are rotated by some angle of order one
  with respect to the basis vectors $(\hat\bX,0)$ and $(0,0,1)$.
\end{enumerate}

To summarize, the longest principal axis is always along $(\hat \bX^\perp,0)$. 
The imaging method is based on this observation.

\subsection{Generalizations}
\label{sect:general}
If the particles have different radii $a_j$ and reflectivities
$\sigma_j$, we can model the cloud using a probability density
function $p(a,\sigma)$ of the joint distribution of the radii and
reflectivities. Then, the previous results hold true, up to the
following minor modification: The covariance function of the scattered
field is of similar form to (\ref{eq:cov2a}), with
\begin{align}
  \label{eq:cov2ahetero}
\EE \big[ u_{\rm s}(\vec\bx_1)\overline{u_{\rm s}}(\vec\bx_2) \big] &=
\frac{k^4 \rho r_o^2e^{-2 k_{\rm d} Z}}{32 \pi |\bX|} \Psi \Big( k
\hat{\bX} \cdot \bx ,\frac{k r_o}{2 |\bX|} \hat{\bX}^\perp
\cdot \bx, k z \Big),
\end{align}
and function
\begin{align}
\Psi ( \chi ,\xi,\zeta)&= \int_0^\infty da\int_0^\infty d \sigma \,
 p(a,\sigma)\left(\sigma \frac{4 \pi a^3}{3}\right)^2 \Psi_{\rm Mie}(
\chi ,\xi,\zeta; ka,\sigma),
    \label{eq:cov2bhetero}
\end{align}
defined by the average of $ \Psi_{\rm Mie}( \chi ,\xi,\zeta; ka,
\sigma)$ given in (\ref{eq:cov2b}). Here we recalled the expression
\eqref{eq:form10} of the constant $\eta$ in \eqref{eq:cov2a}.

We will see in the next section that our imaging algorithm is based entirely on the fact that the decay 
of the correlation function is anisotropic, with very large correlation length in the direction
$(\hat \bX^\perp, 0)$. This strong anisotropy was established in the previous section for all ratios 
$a/\lambda$.
In equation (\ref{eq:cov2bhetero}) we average over $a$, so the conclusion extends to the case of mixtures of particles.

For larger particles that are not spherical, the result should be
qualitatively the same. Indeed, the calculation in appendix
\ref{ap:propsmall} shows that the exponential in equation
\eqref{eq:cov2b}, evaluated at the arguments in \eqref{eq:cov2a}, is
derived independent of the particle size or shape. The kernel ${\cal
  I}_{\rm Mie}$ in \eqref{eq:cov2b} will change for different shapes
of particles, but the relation between the correlation lengths will be
similar to that for spherical particles.

Stronger scattering regimes, that go beyond the Born approximation,
can be modeled using the radiative transport equation \cite{Chandra}
or its simpler, forward scattering version \cite{BorGar}. Such models
of the intensity are more complicated, but as long as the transport
directions are not mixed too much by multiple scattering, which
happens at distances smaller than the transport mean free path, the
result is qualitatively the same.

In summary, imaging based on the qualitative relation between the
correlation lengths displayed in equations~(\ref{eq:correlsmall}-\ref{eq:correlsmallp}) and
\eqref{eq:correlLarge}, which hold in general settings as described
above, is robust to uncertainty of the composition of the cloud of
particles.

\section{Imaging algorithm}
\label{sect:imag}
We now introduce  the algorithm for imaging the axis of the laser beam. We begin in
section \ref{sect:getcorrel} with the estimation of the covariance of
the intensity measured at a camera. Then we recall in section
\ref{sect:levelset} that the level sets of this covariance are
approximate ellipsoids, with axes that depend on the orientation of
the laser. We use this result in section \ref{sect:onegroup} to
extract partial information about the axis of the laser from measurements at a
group of three cameras centered at $\vec{\bX}^{(1)}$, and with mutually
orthogonal planar apertures. In section \ref{sect:twogroup} we show
that it is possible to image the laser  using two such groups of
cameras. The algorithm in section \ref{sect:twogroup} is very
efficient in the two extreme cases $a/\la \ll 1$ and $a/\la \gg 1$,
but it is less efficient when $a/\la \sim 1$. However, the imaging can
be improved if there are three or more groups of cameras, as described
in section \ref{sect:moregroup}. We summarize the imaging algorithm in
section \ref{sect:sumimage}, and also explain there how we quantify
the accuracy of the results.

  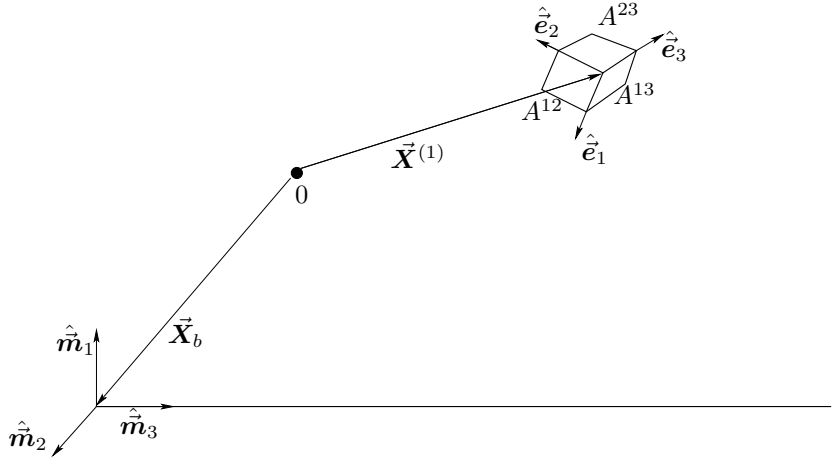
\begin{figure}[t]
\vspace{1.2in}
\begin{center}
\begin{picture}(0,0)%
\hspace{1in}\includegraphics[width = 0.8\textwidth]{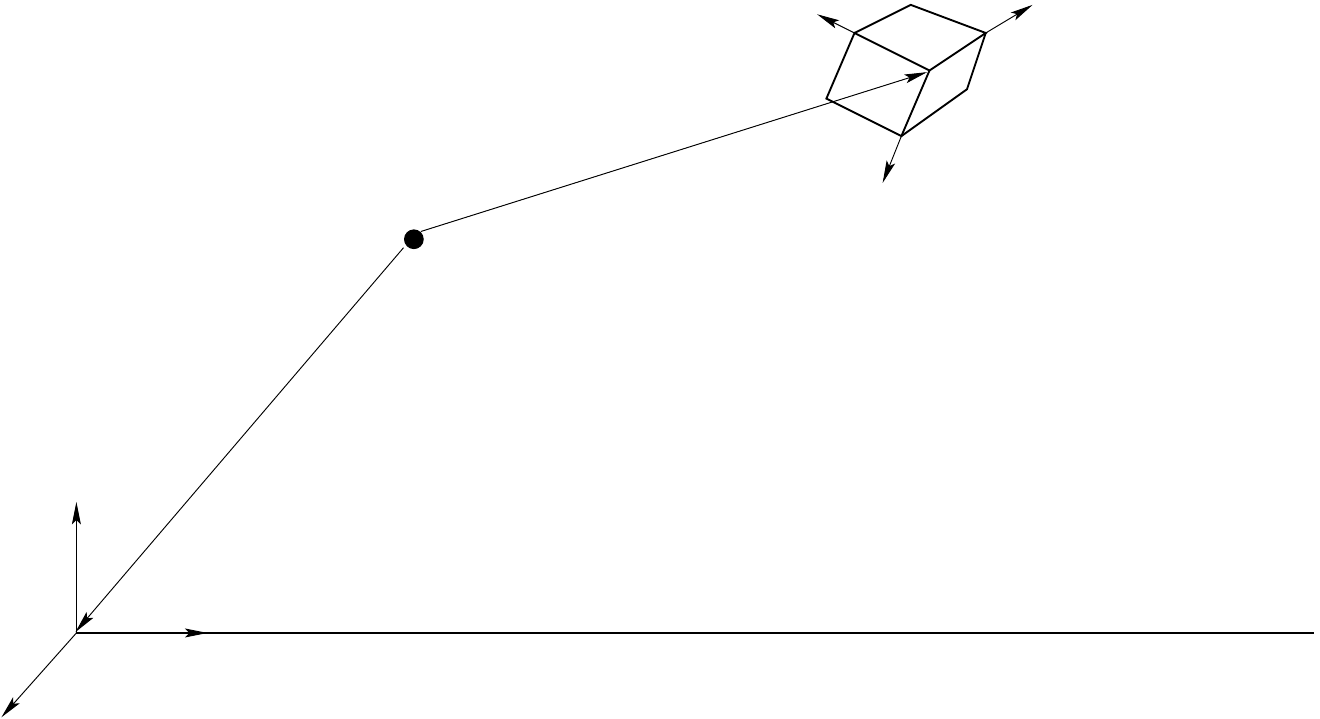}%
\end{picture}%
\setlength{\unitlength}{3947sp}%
\begingroup\makeatletter\ifx\SetFigFont\undefined%
\gdef\SetFigFont#1#2#3#4#5{%
  \reset@font\fontsize{#1}{#2pt}%
  \fontfamily{#3}\fontseries{#4}\fontshape{#5}%
  \selectfont}%
\fi\endgroup%
\begin{picture}(9770,1095)(1568,-1569)
\put(2800,-900){\makebox(0,0)[lb]{\smash{{\SetFigFont{7}{8.4}{\familydefault}{\mddefault}{\updefault}{\color[rgb]{0,0,0}{\normalsize $\hat{\vec{\itbf m}}_1$}}%
}}}}
\put(2500,-1500){\makebox(0,0)[lb]{\smash{{\SetFigFont{7}{8.4}{\familydefault}{\mddefault}{\updefault}{\color[rgb]{0,0,0}{\normalsize $\hat{\vec{\itbf m}}_2$}}%
}}}}
\put(3200,-1420){\makebox(0,0)[lb]{\smash{{\SetFigFont{7}{8.4}{\familydefault}{\mddefault}{\updefault}{\color[rgb]{0,0,0}{\normalsize $\hat{\vec{\itbf m}}_3$}}%
}}}}
\put(4300,30){\makebox(0,0)[lb]{\smash{{\SetFigFont{7}{8.4}{\familydefault}{\mddefault}{\updefault}{\color[rgb]{0,0,0}{\normalsize $0$}}%
}}}}
\put(3500,-850){\makebox(0,0)[lb]{\smash{{\SetFigFont{7}{8.4}{\familydefault}{\mddefault}{\updefault}{\color[rgb]{0,0,0}{\normalsize $\vec\bX_b$}}%
}}}}
\put(4900,250){\makebox(0,0)[lb]{\smash{{\SetFigFont{7}{8.4}{\familydefault}{\mddefault}{\updefault}{\color[rgb]{0,0,0}{\normalsize $\vec\bX^{(1)}$}}%
}}}}
\put(6100,300){\makebox(0,0)[lb]{\smash{{\SetFigFont{7}{8.4}{\familydefault}{\mddefault}{\updefault}{\color[rgb]{0,0,0}{\normalsize $\hat{\vec{\itbf e}}_1$}}%
}}}}
\put(5800,1100){\makebox(0,0)[lb]{\smash{{\SetFigFont{7}{8.4}{\familydefault}{\mddefault}{\updefault}{\color[rgb]{0,0,0}{\normalsize $\hat{\vec{\itbf e}}_2$}}%
}}}}
\put(6600,950){\makebox(0,0)[lb]{\smash{{\SetFigFont{7}{8.4}{\familydefault}{\mddefault}{\updefault}{\color[rgb]{0,0,0}{\normalsize $\hat{\vec{\itbf e}}_3$}}%
}}}}
\put(6200,1150){\makebox(0,0)[lb]{\smash{{\SetFigFont{7}{8.4}{\familydefault}{\mddefault}{\updefault}{\color[rgb]{0,0,0}{\normalsize $A^{23}$}}%
}}}}
\put(6300,650){\makebox(0,0)[lb]{\smash{{\SetFigFont{7}{8.4}{\familydefault}{\mddefault}{\updefault}{\color[rgb]{0,0,0}{\normalsize $A^{13}$}}%
}}}}
\put(5720,550){\makebox(0,0)[lb]{\smash{{\SetFigFont{7}{8.4}{\familydefault}{\mddefault}{\updefault}{\color[rgb]{0,0,0}{\normalsize $A^{12}$}}%
}}}}
\end{picture}%
\end{center}
\caption{Geometric set-up: $\vec\bX^{(1)}$ is the position of the group of cameras, 
$(\hat{\vec{\itbf e}}_1,\hat{\vec{\itbf e}}_2,\hat {\vec {\itbf e}}_3)$ is the orthonormal basis associated with the cameras,
and $A^{12}$, $A^{13}$, $A^{23}$ are the planar apertures of the cameras.
The beam center is at $\vec\bX_b$ and $( \hat{\vec{\itbf m}}_1, \hat{\vec{\itbf m}}_2, \hat{\vec{\itbf m}}_3)$ is the beam orthonormal basis, with $\hat{\vec{\itbf m}}_3$ aligned with the axis of the beam, shown in the figure as the horizontal line.}
\label{fig:frame}
\end{figure}

\subsection{Estimation of the covariance function}
\label{sect:getcorrel}
We consider a group of three cameras centered at $\vec{\bX}^{(1)}$ with mutually
orthogonal\footnote{The apertures do not need to be
  orthogonal, but they should belong to different planes. We choose
  orthogonal planes for convenience.}  planar apertures $A^{12}$, $A^{13}$, $A^{23}$, as in Figure \ref{fig:frame}.
  If we introduce the orthonormal basis $\{\hat{\vec{\itbf e}}_1,\hat
{\vec{\itbf e}}_2,\hat {\vec {\itbf e}}_3\}$ with $\{\hat{\vec{\itbf
    e}}_j,\hat {\vec{\itbf e}}_q\}$ spanning the plane containing  $A^{jq}$ for $1\leq j < q \leq 3$, then  we can define
     explicitly the apertures as the sets
\begin{align}
\label{eq:defAjq}
A^{jq} = \big\{ \vec\bx = \vec\bX^{(1)} + \tilde{x}_j \hat{\vec{\itbf
    e}}_j+\tilde{x}_q \hat {\vec{\itbf e}}_q, \, (\tilde{x}_j,\tilde{x}_q)\in A\big\},\qquad
    A=[0,d_A]^2.
\end{align}

Let us consider one camera, say the one with aperture $A^{12}$, and denote by 
\begin{equation}
  I(\tilde{\bx}) = \big|u_{\rm s}(  \vec\bX^{(1)} + \tilde{x}_1 \hat{\vec{\itbf
    e}}_1+\tilde{x}_2 \hat {\vec{\itbf e}}_1 )\big|^2, \quad \tilde{\bx} =(\tilde{x}_1,\tilde{x}_2)  \in A=[0,d_A]^2,
\label{eq:inv1}
\end{equation}
the measured intensity. 
The empirical covariance function of this intensity is
\begin{equation}
{\mathcal C}(\tilde{\bx}) = \frac{1}{|A_{\tilde{\bx}}|} \int_{A_{\tilde{\bx}}} d \tilde{\bx}'\, I_{\rm c}(\tilde{\bx}')
I_{\rm c}(\tilde{\bx}'+\tilde{\bx}), \qquad A_{\tilde{\bx}} = A \cap (A-\tilde{\bx}), 
\label{eq:invCov}
\end{equation}
for  $\tilde{\bx} \in [-d_A,d_A]^2$, 
where $I_{\rm c}$ is  the centered intensity 
\begin{equation}
  I_{\rm c}(\tilde{\bx}) = I(\tilde{\bx})- \frac{1}{|A|} \int_A d \tilde{\bx}' I(\tilde{\bx}').
\label{eq:inv2}
\end{equation}
Alternatively, we can calculate  the empirical covariance using Fourier
transforms, 
\begin{equation}
{\mathcal C}(\tilde{\bx}) = {\rm FT}^{-1} \big( \big| {\rm FT}( I_{\rm c}) \big|^2
\big)(\tilde{\bx}) ,
\label{eq:inv4}
\end{equation}
where ${\rm FT}$ and ${\rm FT}^{-1}$ denote the modified Fourier transform
and its inverse
\[
{\rm FT}(f)(\tilde{\itbf q}) = \int_A d\tilde{\itbf x} \, f(\tilde{\bx}) e^{i \tilde{\itbf
    q}\cdot\tilde{\itbf x}}, \qquad {\rm FT}^{-1}(\hat{f})(\tilde{\itbf x}) =
\frac{1}{(2\pi)^2 |A|} \int_A d\tilde{\itbf q} \, \hat{f}(\tilde{\itbf q}) e^{-i
  \tilde{\itbf q}\cdot\tilde{\itbf x}}.
\]
In practice, formula \eqref{eq:inv4} can be implemented using the Fast
Fourier Transform (FFT).

Note that for any pair of points $\vec\bX\pm \vec\bx/2$ in $A^{12}$, the statistical covariance function and
the empirirical covariance function are approximately the same
\begin{equation}
  {\rm Cov}(|u_{\rm s}(\vec \bX + \vec \bx/2)|^2, |u_{\rm s}(\vec \bX - \vec
  \bx/2)|^2) \approx {\mathcal C}(\tilde{\bx}), \qquad \tilde{\bx} = (\vec\bx \cdot\hat
       {\vec{\itbf e}}_1, \vec\bx \cdot\hat {\vec{\itbf e}}_2),
  \label{eq:inv3}
\end{equation}
provided the area $|A|$ of the camera is large compared to the area of a speckle spot.
The empirical correlation function of the intensity is 
\begin{equation}
%{\rm Corr} \big( |u_{\rm s}(\vec\bX + \vec \bx/2)|^2,|{u_{\rm s}}(\vec\bX -
%\vec \bx/2)|^2\big) \approx 
\mathscr{C}(\tilde{\bx}) =
\frac{ \mathcal{C}(\tilde{\bx}) }{\mathcal{C}((0,0))} ,
\label{eq:CorrelationFunc}
\end{equation}
and we note that with one camera we can only evaluate the correlation
function in the plane of its aperture. To estimate the correlation function for
all $\vec \bx \in \mathbb{R}^3$, we need a group of cameras centered
at $\vec \bX^{(1)}$, with apertures lying in different planes, as explained
in section \ref{sect:onegroup}.

\subsection{The level sets of the correlation function}
\label{sect:levelset}
In this section we consider the level sets of the statistical correlation
function of the intensity at values close to one, which can be approximated
by ellipsoids as shown in section \ref{sect:level}. 
We describe the  axes of this ellipsoid in
a general set up, for an arbitrary orientation of the axis of the laser beam.

It is convenient to introduce a new system of
coordinates with orthonormal "beam basis"
$\{\hat{\vec{\itbf m}}_1,\hat{\vec{\itbf m}}_2,\hat{\vec{\itbf
    m}}_3\}$. We call it the beam basis because it is defined
relative to the axis of the
laser beam, the line $\{ \vec\bX_b+s \hat{\vec{\itbf Y}}_b, \, s\in \RR\}$
along the unit vector $\hat{\vec{\itbf Y}}_b$, parametrized by the
arc-length~$s$.  The origin of the arc-length is arbitrary, so
$\vec\bX_b$ can be any point on the axis.  Note that 
the beam basis also depends on the center $\vec \bX^{(1)}$ of the camera,
which lies, as the axis of the laser, in the plane
spanned by the vectors $\vec \bX^{(1)} - \vec \bX_b$ and $\hat{\vec{\itbf
    Y}}_b$.  We define the beam basis by
\begin{align}
  \hat{\vec{\itbf m}}_3 = \hat{\vec{\itbf Y}}_b,\qquad
  \hat{\vec{\itbf m}}_2 = \frac{\hat{\vec{\itbf m}}_3 \times
    (\vec\bX^{(1)}-\vec\bX_b)}{\| \hat{\vec{\itbf m}}_3 \times
    (\vec\bX^{(1)}-\vec\bX_b) \|}, \qquad \hat{\vec{\itbf m}}_1 &=
  \hat{\vec{\itbf m}}_2 \times \hat{\vec{\itbf m}}_3 \label{eq:m1},
\end{align}
and note that in section \ref{sect:stat} we considered the special case $\vec \bX_b
= (0,0,0)$ and $\hat{\vec{\itbf Y}}_b=(0,0,1)$, so that 
 $\hat{\vec{\itbf m}}_1 = (\hat{\bX}^{(1)},0)$
and $\hat{\vec{\itbf m}}_2 = (\hat{\bX}^{(1),\perp},0)$.  The 
basis \eqref{eq:m1} is defined for an arbitray
orientation of the axis of the beam and origin of coordinates, and it is unknown in imaging. 
We only know the basis
$\{\hat{\vec{\itbf e}}_1,\hat {\vec{\itbf e}}_2,\hat {\vec {\itbf
    e}}_3\}$ defined relative to the group of  cameras.

To write explicitly the correlation function of the intensity at two
points $\vec \bX \pm \vec \bx/2$ in $A^{12}\cup A^{13}\cup A^{23}$, we decompose the offset
vector $\vec \bx$ in the beam basis
\begin{equation}
  \vec\bx = \sum_{j=1}^3 x_j \hat{\vec{\itbf m}}_j,
  \label{eq:defdecompp}
\end{equation}
and scale its components by the characteristic correlation lengths
described in section \ref{sect:stat},
\begin{equation}
x_1 = \frac{1}{k} \chi, \qquad x_2 = \frac{2|\bX^{(1)}-\bX_b|}{k r_o} \xi,
\qquad x_3 = \frac{1}{k} \zeta.
\label{eq:defcomp}
\end{equation}
Note that the transverse distance $|\bX|$ is now $|\bX^{(1)}-\bX_b|$,
with 
$$
  \vec\bX^{(1)} = \sum_{j=1}^3 X^{(1)}_j \hat{\vec{\itbf m}}_j,\qquad
  \vec\bX_b = \sum_{j=1}^3 X_{b,j} \hat{\vec{\itbf m}}_j,\qquad
  \bX^{(1)}-\bX_b = \sum_{j=1}^2 ( X^{(1)}_j - X_{b,j} ) \hat{\vec{\itbf m}}_j .
$$
This comes from the following lemma, that states that the dependence of the statistical covariance function with 
respect to the mid point
$\vec\bX$ is so slow that we can replace $\vec{\bX}$ by $\vec\bX^{(1)}-\vec\bX_b$, with negligible error.

\vspace{0.05in}
\begin{lemma}
\label{lem:replaceX}
Under the scaling assumption $d_A \ll r_o$ stated in \eqref{eq:form12}, and for any $\vec\bx_1,\vec\bx_2 \in A^{12} \cup A^{13}\cup A^{23}$, 
Propositions \ref{prop:small}-\ref{prop:large} hold true with $\vec\bx=\vec\bx_1-\vec\bx_2$, $\vec\bX$ replaced by 
$\vec{\bX}_c=\vec\bX^{(1)}-\vec\bX_b$,
$\bX$ replaced by $\bX_c= \bX^{(1)}-\bX_b$, and  the unit vector 
$\hat{\bX}$ replaced by $\hat{\bX}_c=   (\bX^{(1)}-\bX_b)/ |  \bX^{(1)}-\bX_b|$.
\end{lemma}

\vspace{0.05in}
\noindent
\begin{proof}
We need to check that, for any $|\vec\bx| \leq \lambda |\bX_c| / r_o$, the arguments of the functions $\Psi$
in the propositions do not change at order one when  the mid point $\vec\bX$ is replaced by  $\vec\bX_c$,
$\bX$ is replaced by $\bX_c$, and $\hat{\bX}$ is replaced by $\hat{\bX}_c$. This follows 
from the estimates
\begin{align*}
\big|k \hat{\bX}\cdot \bx -k \hat{\bX}_c\cdot \bx \big| &\approx \frac{k}{|\bX_c|} \Big| \bx \cdot \big\{(\bX - \bX_c) - 
 \hat{\bX}_c \big[ \hat{\bX}_c \cdot
(\bX - \bX_c)\big]\big\} \Big| \\
&\lesssim \frac{d_A}{r_o} \ll 1 ,
\end{align*}
and 
\begin{align*}
\Big| \frac{kr_o}{2|\bX|}  \hat{\bX}^\perp\cdot \bx -\frac{kr_o}{2|\bX_c|}  \hat{\bX}_c^{\perp}\cdot \bx \Big|
\approx &
\frac{k r_o}{2|\bX_c|^2} \Big| 2\big(\hat \bX_c \cdot \bx^\perp\big) \big[\hat \bX_c \cdot 
(\bX-\bX_c)\big] \\
 &-\bx^\perp \cdot (\bX - \bX_c) \Big|
 \lesssim  \frac{d_A}{|\bX_c|} \sim \frac{d_A}{l_{\bx}} \ll 1 ,
 \end{align*}
where the superscript $\perp$ denotes  rotation of the  vectors  $\hat{\bX}$, $\hat{\bX}_c$, and $\bx$ by ninety degrees, in the cross-range plane $(\hat{\vec{\itbf m}}_1,\hat{\vec{\itbf m}}_2)$. \end{proof}

Therefore, the expression of the correlation function of the intensity is still (\ref{eq:LCorrelMod}) in terms of $\chi$, $\xi$, and $\zeta$
defined by (\ref{eq:defcomp}), 
and the level set ${S}_{\mathfrak{L}}$ of the correlation
function at value $1 - {\mathfrak L}$, for $0 < {\mathfrak L} \ll
1$ is the ellipsoid
\begin{equation}
 x_1^2 \left(\frac{k^2H_{11}}{\mathfrak{L}} \right) + x_3^2 \left(\frac{k^2 H_{33}}{\mathfrak{L}}\right)+2 x_1 x_3
  \left(\frac{k^2 H_{13}}{\mathfrak{L}} \right) + x_2^2 \left[\Big(\frac{k r_o}{|\bX^{(1)} -\bX_b|}\Big)^2
    \frac{H_{22}}{{\mathfrak L}}\right] = 1,
\label{eq:ellipse}
\end{equation}
in terms of $x_1$, $x_2$, and $x_3$ defined by (\ref{eq:defdecompp}).
One principal axis of the ellipsoid is along the unit vector $\hat{\vec{\itbf m}}_2$
and the other principal axes are in the plane containing the center of the
camera and the axis of the laser beam, spanned by $\hat{\vec{\itbf
    m}}_1$ and $\hat{\vec{\itbf m}}_3$. 
As in section \ref{sect:level}, the main observation is that
 the ellipsoid \eqref{eq:ellipse} has the longest principal axis along $\hat{\vec{\itbf m}}_2$.

\subsection{Estimation with one group of cameras}
\label{sect:onegroup}
We now explain how to use a group of three cameras centered at $\vec
\bX^{(1)}$ to estimate the ellipsoids that approximate the level sets of the
correlation function of the intensity. We can then extract information
about the axis of the laser beam using the relations between the principal axes of
the ellipsoids and the beam basis described in the previous
section.

To determine the correlation function ${\rm Corr} \big( |u_{\rm s}(\vec\bX
+\vec \bx/2)|^2,|{u_{\rm s}}(\vec\bX-\vec \bx/2)|^2\big)$ for all vectors
$\vec \bx \in \RR^3$, we use the three cameras centered at $\vec\bX^{(1)}$
 with apertures $A^{jq}$ defined in \eqref{eq:defAjq}, for $1 \leq j<q \leq 3$.
 
As shown in the previous section, the correlation function as a function of $\vec\bx$ 
 can be approximated by a Gaussian near its peak at ${\bf 0}$. This Gaussian can be
represented by a symmetric and positive definite matrix ${\bf U}\in
\RR^{3 \times 3}$, with normalized eigenvectors $(\hat{\vec
  \bu}_j)_{j=1,2,3}$ that are along the principal axes of the
ellipsoids, the level sets. The eigenvalues of ${\bf U}$ equal the
lengths of these axes raised to the power $-2$.

Let $U_{jq} = \hat{\vec {\itbf e}}_j \cdot {\bf U} \hat{ \vec {\itbf
    e}}_q $ be the components of ${\bf U}$ in the known basis 
$\{ \vec{\itbf e}_1,\vec{\itbf e}_2,\vec{\itbf e}_3\}$    
 and denote
\begin{equation}
\label{eq:defUjq}
\boldsymbol{\Pi}_{jq} {\bf U} = \begin{pmatrix} U_{jj} & U_{jq}\\
  U_{jq} & U_{qq}, 
  \end{pmatrix}, \qquad 1 \leq j<q \leq 3.
\end{equation}
Let also
$\mathscr{C}^{jq}(\tilde{\bx})$ denote the correlation function
\eqref{eq:CorrelationFunc} obtained with the camera with aperture
$A^{jq}$. We estimate the matrix \eqref{eq:defUjq} by the minimizer
\begin{equation}
  {\bf V}^{jq} = \dessous{\rm argmin}{{\bf V} \in \RR^{2\times 2}}
       {\cal E}^{jq}({\bf V}) \mbox{ subject to ${\bf V}^t={\bf
           V}$ and ${\bf V}\geq 0$,} \label{eq:optimize}
\end{equation}
of the objective function
\begin{equation}
{\cal E}^{jq}({\bf V}) = \int_{[-d_A,d_A]^2} d \tilde{\bx} \,
|\mathscr{C}^{jq}(\tilde{\bx}) - {\cal G}(\tilde{\bx},{\bf V})|^2 {\bf
  1}_{\mathscr{C}^{jq}(\tilde{\bx}) >1-\mathfrak{L}},
\end{equation}
where the correlation function is fitted by the Gaussian
\begin{equation}
  {\cal G}(\tilde{\bx},{\bf V}) = \exp
\big\{ [\ln (1-\mathfrak{L})] \tilde{\bx}^t {\bf V} \tilde{\bx} \big\},
\end{equation}
at points in the level sets of value greater than
$1-\mathfrak{L}$. The value $\mathfrak{L}$, chosen by the user, should
be small and positive.

In practice, due to measurement errors and imprecise solutions of \eqref{eq:optimize}, the minimizers ${\bf
  V}^{jq}$ give different estimates of $U_{jj}$, for $1\leq j < q \leq 3$. Thus, we incorporate all the results in
another optimization problem 
\begin{equation}
{\boldsymbol{\mathfrak{U}}} = \dessous{\rm argmin}{{\bf V} \in \RR^{3
    \times 3}} \sum_{1 \le j < q \le 3} \| \boldsymbol{\Pi}_{jq}
{\bf V} -{\bf V}^{jq} \|^2,
\label{eq:finopt}
\end{equation}
where $\| \cdot \|$ is the Frobenius norm, and estimate the matrix
${\bf U}$ by the minimizer ${\boldsymbol{\mathfrak{U}}}$. This is a
symmetric matrix with entries
\begin{align}
 {\mathfrak U}_{11} &= \frac{V_{11}^{12}+V_{11}^{13}}{2}, \quad
 \quad {\mathfrak U}_{22} = \frac{V_{22}^{12}+V_{11}^{23}}{2},
 \quad \quad {\mathfrak U}_{33} = \frac{ V_{22}^{13} +
   V_{22}^{23}}{2} , \nonumber \\ {\mathfrak U}_{12}&=
 {\mathfrak U}_{21} = V^{12}_{12}, \quad \quad
 {\mathfrak U}_{13}= {\mathfrak U}_{31}= V^{13}_{12},
 \quad \quad {\mathfrak U}_{23}={\mathfrak U}_{32} =
 V^{23}_{12}. \label{eq:estUmat}
\end{align}
It has positive trace, equal to the average of the traces of the positive
definite matrices ${\bf V}^{jq}$, so  the largest
eigenvalue of ${\boldsymbol{\mathfrak{U}}}$ is positive.  We know from
the discussion in the previous section that ${\bf U}$ has at least
one small eigenvalue, corresponding to the eigenvector along
$\hat{\vec {\itbf m}}_2$.  We also know from Weyl's theorem
\cite{parlett} that the eigenvalues of ${\boldsymbol{\mathfrak{U}}}$
are within the distance $\|{\boldsymbol{\mathfrak{U}}} - {\bf   U}\|_2$ 
of those of ${\bf U}$. Thus, ${\boldsymbol{\mathfrak{U}}}$
may have a negative eigenvalue, with small absolute value determined
by measurement errors.

The orthonormal eigenvectors $(\hu_j)_{j=1,2,3}$ of
${\boldsymbol{\mathfrak{U}}}$ approximate the principal axes of the
ellipsoid, which are aligned with the eigenvectors $(\hat{\vec
  \bu}_j)_{j=1,2,3}$ of the exact matrix ${\bf U}$. The accuracy of
the approximation depends on the sensitivity of the eigenvectors to
measurement errors, which depends in turn on the gap between the
eigenvalues. The more robust eigenvectors correspond to eigenvalues
that are well separated from the rest \cite{parlett}, so we base our
imaging on them. The discussion in the previous section shows that,
depending on the size of the particles, we have three cases:

\vspace{0.03in}
\begin{enumerate}
  \itemsep 0.03in
\item For small particles with radius $a$ satisfying $k a\ll 1$, the
  matrix ${\bf U}$ has two large eigenvalues of the same order, and
  a much smaller third eigenvalue, by a factor of $(r_o/|\bX^{(1)}-\bX_b|)^2 \ll
  1$. Because this third eigenvalue is well separated from the larger
  ones, the corresponding eigenvector
  $\hu_3$ is robust to measurement
  errors, and we use it to approximate $\hat{\vec \bu}_3 =
  \hat{\vec{\itbf m}}_2$. 
\item For large particles with radius $a$ satisfying $ka \gg 1$, the
  matrix ${\bf U}$ has one large eigenvalue corresponding to the
  eigenvector $\hat{\vec{\itbf u}}_1 \approx \hat{\vec{\itbf m}}_1$
  and two much smaller eigenvalues. Here, the more robust eigenvector
  is $\hu_1$, and we use it to
  approximate $\hat{\vec{\itbf m}}_1$.
\item For particles of intermediate size, the matrix ${\bf U}$ has
  three distinct eigenvalues, with separation that depends on the
  scattering kernel $\mathfrak{S}$. If the gap between the
  second and third eigenvalues of the estimated matrix
  ${\boldsymbol{\mathfrak{U}}}$ is small, we use
  $\hu_1$ to approximate
  $\hat{\vec{\itbf u}}_1$. This vector is no longer aligned with
  $\hat{\vec{\itbf m}}_1$, but it lies in the plane containing the
  center of the camera and the axis of the laser beam, spanned by
  $\hat{\vec{\itbf m}}_1$ and $\hat{\vec{\itbf m}}_3$. Otherwise, if
  the third eigenvalue of ${\boldsymbol{\mathfrak{U}}}$ is well
  separated from the larger ones, we approximate $\hat{\vec{\itbf
      m}}_2$ by $\hu_3$.
\end{enumerate}

\subsection{Imaging with two groups of cameras}
\label{sect:twogroup}
The results of the previous section show that depending on the cloud
of particles, we may have three scenarios:
\\
\textbf{Scenario 1:} Where we can estimate
the unit vector $\hat{\vec{\itbf m}}_2$ normal  to the plane containing the center
of the camera and the axis of the laser, using the eigenvector $\hu_3$
corresponding to the smallest eigenvalue of
${\boldsymbol{\mathfrak{U}}}$. \\
\textbf{Scenario 2:} Where the eigenvector $\hu_3$ is
too sensitive to measurement errors, but we can estimate the 
vector $\hat{\vec{\itbf m}}_1$ using the eigenvector $\hu_1$ of
${\boldsymbol{\mathfrak{U}}}$, for the largest eigenvalue. This occurs
for large particles. \\
\textbf{Scenario 3:} Where the only robust eigenvector is $\hu_1$,
but its direction is not close to that of the vector
$\hat{\vec{\itbf m}}_1$. This occurs for particles of intermediate
size.

\vspace{0.03in} \noindent In the first two scenarios, we can image the laser
beam using two groups of cameras, as we explain in this section. The
last scenario requires more measurements, and is discussed in the next
section. 

We use henceforth the notation $\vec\bX^{(j)}$ for the
centers of the groups of cameras, with $j\ge 1$, and assume that the
laser beam axis and any two of these centers do not lie in the same
plane.  We also let $\big(\hu_q^{(j)}\big)_{q=1,2,3}$ be the
eigenvectors of the estimated matrix
${\boldsymbol{\mathfrak{U}}}^{(j)}$ with the $j$-th group of cameras.
To distinguish the exact laser beam axis $\{ \vec\bX_b+s
\hat{\vec{\itbf Y}}_b, \, s\in \RR\}$ from the estimated one, we index
the latter by a star, as in $\{ \vec\bX^\star_b+s \hat{\vec{\itbf
    Y}}_b^\star, \, s\in \RR\}$.
\vspace{0.05in}
\begin{alg}
\label{alg:1} This algorithm applies to
scenario 1, and uses as inputs $\vec\bX^{(j)}$ and $\hu_3^{(j)}$, for
$j = 1,2$.  The output is the estimated laser beam axis $\{
\vec\bX^\star_b+s \hat{\vec{\itbf Y}}_b^\star, \, s\in \RR\}$, with
\begin{equation}
  \hat{\vec{\itbf Y}}_b^\star = \frac{\hu_3^{(1)} \times \hu_3^{(2)}}{
    |\hu_3^{(1)} \times \hu_3^{(2)}|}, \qquad \vec\bX_b^\star = c_1
  \hu_3^{(1)} + c_2 \hu_3^{(2)},
  \label{eq:Alg1}
\end{equation}
and coefficients
\begin{align}
  c_1 &= \frac{ \hu_3^{(1)}\cdot \vec\bX^{(1)} - \big[\hu_3^{(1)}\cdot
      \hu_3^{(2)}\big] \hu_3^{(2)} \cdot
    \vec\bX^{(2)}}{1-\big[\hu_3^{(1)}\cdot \hu_3^{(2)}\big]^2}
  ,\nonumber \\ c_2 &= \frac{ \hu_3^{(2)}\cdot \vec\bX^{(2)} -
    \big[\hu_3^{(1)}\cdot \hu_3^{(2)}\big] \hu_3^{(1)} \cdot
    \vec\bX^{(1)}}{1-\big[\hu_3^{(1)}\cdot
      \hu_3^{(2)}\big]^2}. \label{eq:Alg2}
\end{align}
\end{alg}

In scenario 1, the vectors $\hu_3^{(j)}$ approximate the unit vectors normal to
the two planes defined by the axis of the beam and the centers
$\vec\bX^{(j)}$ of the two groups of cameras. These normal vectors  are not collinear,
because these two planes do not coincide,  so the laser axis must be collinear with their
cross-product, as stated in \eqref{eq:Alg1}. We also have that
$\vec{\bX}^{(j)}-\vec\bX_b$ must be orthogonal to $\hu_3^{(j)}$, for
$j = 1,2,$ and seek $\vec\bX_b$ in the plane orthogonal to the 
laser axis. Thus, we represent $\vec\bX_b^\star$ in \eqref{eq:Alg1} as a
vector in the span of $\hu_3^{(1)}$ and $\hu_3^{(2)}$, and obtain the
expression \eqref{eq:Alg2} of the coefficients $c_1$ and $c_2$ by
solving the linear system of equations
\[
(\vec{\bX}^{(j)}-\vec\bX_b^\star)\cdot \hu_3^{(j)} = 0, \quad j = 1,2.
\]

\vspace{0.05in}
\begin{alg}
  \label{alg:2}
This algorithm applies to scenario 2, and uses as inputs
$\vec\bX^{(j)}$ and $\hu_1^{(j)}$, for $j = 1,2$.  The output is the
estimated laser beam axis $\{ \vec\bX_b^\star+s \hat{\vec{\itbf Y}}_b^\star, \, s\in
\RR\}$, with 
\begin{equation}
  \hat{\vec{\itbf Y}}_b^\star = \frac{\hu_1^{(1)} \times \hu_1^{(2)}}{
    |\hu_1^{(1)} \times \hu_1^{(2)}|}, \qquad \vec\bX_b^\star = c_1
  \hu_1^{(1)} + c_2 \hu_1^{(2)},
  \label{eq:Alg3}
\end{equation}
and coefficients 
\begin{align}
  c_1 &= \frac{\vec\bX^{(2)} \cdot \big\{ \hu_1^{(1)} - \big[
      \hu_1^{(1)}\cdot \hu_1^{(2)}\big] \hu_1^{(2)}\big\}}{1 -
    \big[\hu_1^{(1)}\cdot \hu_1^{(2)}\big]^2} ,\nonumber \\ c_2 &=
\frac{\vec\bX^{(1)} \cdot \big\{ \hu_1^{(2)} - \big[
      \hu_1^{(1)}\cdot \hu_1^{(2)}\big] \hu_1^{(1)}\big\}}{1 -
    \big[\hu_1^{(1)}\cdot \hu_1^{(2)}\big]^2}. \label{eq:Alg4}
\end{align}
\end{alg}

In scenario 2, the vectors $\hu_1^{(j)}$ approximate the unit vectors normal
to the axis of the laser beam, in the planes defined by this axis and
the centers $\vec\bX^{(j)}$ of the two groups of cameras, for $j =
1,2$. These vectors are not collinear, because the two planes do not
coincide, so their cross-product defines the orientation of the axis
of the laser, as in \eqref{eq:Alg3}. The expression of
$\vec\bX_b^\star$ in \eqref{eq:Alg3} states that it is a vector in the
plane orthogonal to the axis of the laser, spanned by $\hu_1^{(1)}$ and
$\hu_1^{(2)}$.  To determine the coefficients in \eqref{eq:Alg4}, we
use that
\[
\vec \bX^{(j)} - \vec \bX_b^\star \in \mbox{span}\{\hu_1^{(j)},\hu_1^{(1)}
\times \hu_2^{(2)} \}, \quad j = 1, 2.
\]
Equivalently,
\[
\vec \bX^{(j)} - \vec \bX_b^\star \perp \hu_1^{(j)} \times
\big[\hu_1^{(1)} \times \hu_2^{(2)} \big], \quad j = 1, 2,
\]
and substituting \eqref{eq:Alg3} in these equations we obtain a linear
system for the coefficients $c_1$ and $c_2$. The solution of this
system is \eqref{eq:Alg4}.

\subsection{Imaging with three or more groups of cameras}
\label{sect:moregroup}
Algorithm \ref{alg:1} fails in scenario~3, because the matrices
${\boldsymbol{\mathfrak{U}}}^{(j)}$ have two very small eigenvalues,
and the eigenvectors $\hu_3^{(j)}$ are not robust to measurement
errors. Thus, imaging must be based on the leading eigenvectors
$\hu_1^{(j)}$. Algorithm \ref{alg:2} uses these eigenvectors, but its
output is not a good approximation of the axis of the laser, because
the vectors $\hu_1^{(j)}$ are not orthogonal to this axis. They are
rotated by an angle that is unknown and can only be estimated using
knowledge of the scattering properties of the cloud (recall the last
case in section \ref{sect:onegroup}). We assume no such knowledge, so
in scenario 3 we cannot image well using two groups of cameras. In
this section we show how to improve the results using more
measurements, at $N_c \ge 3$  groups of cameras.

The basic idea of the algorithm is that, if we had a point $\vec
\bX^\star$ on the axis of the laser, so that $\vec\bX^{(j)} - \vec \bX^\star$
is not collinear to $\hu_{1}^{(j)}$, then we could approximate well
the unit vector normal to the plane containing $\vec\bX^{(j)}$ and the axis of the
laser, i.e., approximate the basis vector
\[
  \hat{\vec{\itbf m}}_2^{(j)} \approx \frac{\big(\vec{\bX}^{(j)} -
    \vec{\bX}^\star\big) \times \hu_1^{(j)}}{\|\big(\vec{\bX}^{(j)}
    - \vec{\bX}^\star\big) \times \hu_1^{(j)}\|}.
\]
We do not know $\vec \bX^\star$, we only have its  estimate \eqref{eq:Alg4}
obtained with two groups of cameras,  and this will likely lie off the axis of the
laser. However, we can search for $\vec\bX^\star$, such that the
vectors
\[
\hat{\vec{\itbf w}}^{(j)} :=\frac{\big(\vec{\bX}^{(j)} -
    \vec{\bX}^\star\big) \times \hu_1^{(j)}}{\|\big(\vec{\bX}^{(j)}
    - \vec{\bX}^\star\big) \times \hu_1^{(j)}\|},  
\quad j = 1, \ldots, N_c \ge 3, 
\]
lie in a two-dimensional space, which is the plane orthogonal to the
axis of the laser.

\vspace{0.05in}
\begin{alg}
  \label{alg:3}
  The inputs are: the centers $\vec\bX^{(j)}$ of $N_c$
  groups of cameras, the eigenvectors $\hu_1^{(j)}$, for $j =
  1, \ldots, N_c$, and the initial guess $\vec\bX_0$ of $\vec
  \bX^\star$ calculated using the second equation in \eqref{eq:Alg3},
  with coefficients \eqref{eq:Alg4}.  The output is the estimated
  laser beam $\{\vec \bX_b^\star + s \hat{\vec{\itbf Y}}_b^\star, ~ s
  \in \RR\}$, obtained using the following steps: \\ \textbf{Step 1:}
  Search for $\vec\bX^\star$ in the plane defined by $\vec\bX^{(j)}$,
  with $j = 1, 2, 3$. Parametrize the search point by
  \[
  \vec \bX_{{\itbf t}} = t_1 \vec{\btheta}_1 + t_2 \vec{\btheta}_2 +
  \vec \bX_0, \qquad {\itbf t} = (t_1,t_2),
  \]
  where $ \vec{\btheta}_1$ and $\vec{\btheta}_2$ are the left singular
  vectors of the $3 \times 2$ matrix $\big( \vec{\bX}^{(2)}- \vec{\bX}^{(1)},
  \vec{\bX}^{(3)}- \vec{\bX}^{(1)} \big)$. The initial guess corresponds to
  ${\itbf t} = (0,0).$ \\ \textbf{Step 2:} Search for the optimal
  ${\itbf t}^\star$ and set $\vec \bX^\star = \vec \bX_{{\itbf t}^\star}$.
  The optimal ${\itbf t}^\star$ is the minimizer of the objective function
  \[
   \mathcal{O}({\itbf t}) = \log \frac{\Sigma(3)}{\Sigma(2)},
  \]
  where  $(\Sigma(1),\Sigma(2),\Sigma(3))$ are the singular values
  of 
the $3\times N_{\rm c}$ matrix $\big( \hat{\vec{\itbf w}}^{(1)}_{{\itbf t}}, \ldots,
\hat{\vec{\itbf w}}^{(N_c)}_{{\itbf t}} \big)$, sorted in decreasing
order.  The columns of this matrix are the unit vectors
\[
\hat{\vec{\itbf w}}^{(j)}_{{\itbf t}} :=\frac{\big(\vec{\bX}^{(j)} -
    \vec{\bX}_{\itbf t}\big) \times \hu_1^{(j)}}{\|\big(\vec{\bX}^{(j)}
    - \vec{\bX}_{\itbf t}\big) \times \hu_1^{(j)}\|},  
\quad j = 1, \ldots, N_c. 
\]
\textbf{Step 3:} Estimate $\hat{\vec{\itbf Y}}_{b}^\star$ as the third left
singular vector of $\big( \hat{\vec{\itbf w}}^{(1)}_{{\itbf t}}, \ldots,
\hat{\vec{\itbf w}}^{(N_c)}_{{\itbf t}} \big)$, corresponding to the near zero
singular value, per the optimization at step 2.\\ \textbf{Step 4:}
Estimate $\hat\bX_b^\star$ using the second equation in \eqref{eq:Alg1},
with coefficients \eqref{eq:Alg2} and $\hu_3^{(j)}$ replaced by
$\hat{\vec{\itbf w}}^{(j)}_{{\itbf t}^\star}$ and $j = 1,2.$
\end{alg}

The parametrization at step 1 of this algorithm reduces the search
space to two dimensions, in the plane defined by the centers of the
first three groups of cameras. It assumes that this plane is not
collinear to the axis of the laser, which is generally the case.  In
principle, if there are more than three cameras, the search may be
done in the plane defined by any three of them, and the results can be
compared for consistency.

Note that by the ordering of the singular values of the matrix $\big( \hat{\vec{\itbf w}}^{(1)}_{{\itbf t}}, \ldots,
\hat{\vec{\itbf w}}^{(N_c)}_{{\itbf t}} \big)$, we have $\Sigma(3)/\Sigma(2) \le 1$, so
the objective function $\mathcal{O}({\itbf t})$ is negative valued. At
the optimal point ${\itbf t} = {\itbf t}^\star$, the range of this
matrix should be approximately the plane orthogonal to the axis of the
laser. Thus, we expect $\Sigma(1) \sim \Sigma(2) \gg \Sigma(3)
\approx 0$, which motivates the definition of the objective function.

\subsection{The imaging algorithm and quantification of estimation errors}
\label{sect:sumimage}
We begin with the summary of  the imaging algorithm:

\vspace{0.05in}
\begin{alg}
  \label{alg:4}
  The inputs are: The centers $\{\vec \bX^{(j)}\}_{j=1,\ldots, N_c}$
  of $N_c$ groups of cameras, and the estimated correlation function
  for each of them, calculated as explained in section
  \ref{sect:getcorrel}. The output is the estimated axis of the laser beam
  $\{\vec \bX_b^\star + s \hat{\vec{\itbf Y}}_b^\star, ~ s \in \RR\}$ obtained
  using the following two steps: 
  
  \vspace{0.03in}
  \textbf{Step 1:} Estimate the
  matrices ${\boldsymbol{\mathfrak{U}}}^{(j)}$, for $j = 1, \ldots,
  N_c$, as explained in section \ref{sect:onegroup}.

  \vspace{0.03in}
 \textbf{Step
    2:} There are two cases: 
    
    \vspace{0.03in}
    \begin{enumerate} \item If the smallest
    eigenvalue of ${\boldsymbol{\mathfrak{U}}}^{(j)}$ is well
    separated from the others, use the eigenvector
    $\hu_3^{(j)}$ to estimate the unit vector normal to the plane containing $\vec
    \bX^{(j)}$ and the axis of the laser, for $j = 1, 2$. Then estimate
    this  axis using Algorithm \ref{alg:1} and stop.
  \item Otherwise, use the leading eigenvectors $\hu_1^{(j)}$, for $j
    = 1, \ldots, N_c$ to image as follows:
    \begin{enumerate} \item[(i)] If $N_c = 2$, estimate the axis of the laser beam using
      Algorithm \ref{alg:2}.
      \item[(ii)] If $N_c \ge 3$, estimate the
        axis of the laser beam using Algorithm \ref{alg:3}.
      \end{enumerate}
    \end{enumerate}
\end{alg}

\vspace{0.05in} It remains to quantify the estimation error.  Recall
that the true laser axis is the line $\{ \vec\bX_b+s \hat{\vec{\itbf
    Y}}_b, \, s\in \RR\}$. We compare it with the estimated axis $\{
\vec\bX_b^\star+s \hat{\vec{\itbf Y}}_b^\star, \, s\in \RR\}$ using
two quantifiers that are independent on the parametrization of these
lines, which is arbitrary. The first quantifier is the angle $\beta$
between the unit vectors $\hat{\vec{\itbf Y}}_b$ and $\hat{\vec{\itbf
    Y}}_b^\star$:
\begin{equation}
\label{def:angle}
\beta = \mbox{arccos}\big( \big|\hat{\vec{\itbf Y}}_b\cdot \hat{\vec{\itbf
    Y}}_b^\star \big|\big) ,
\end{equation}
which gives the error in the estimated orientation of the laser beam. Here we
take absolute values because the same line is defined by both
$\hat{\vec{\itbf Y}}_b$ and $-\hat{\vec{\itbf Y}}_b$. The second
quantifier is the distance between the two lines (the true beam axis and the estimated one):
\begin{equation}
\label{def:distance}
d = \min_{s,s'\in \RR} \big\| \vec\bX_b+s\hat{\vec{\itbf Y}}_b -
\vec\bX_b^\star - s' \hat{\vec{\itbf Y}}_b^\star \big\| .
\end{equation}
The minimizers in this equation are  
\begin{eqnarray*}
s = \frac{[ - \hat{\vec{\itbf Y}}_b + (\hat{\vec{\itbf Y}}_b\cdot
    \hat{\vec{\itbf Y}}_b^\star) \hat{\vec{\itbf Y}}_b^\star ] \cdot
  (\vec\bX_b^\star-\vec\bX_b)}{1-(\hat{\vec{\itbf Y}}_b\cdot \hat{\vec{\itbf
      Y}}_b^\star)^2} , \quad  s'= \frac{[\hat{\vec{\itbf Y}}_b^\star -
    (\hat{\vec{\itbf Y}}_b\cdot \hat{\vec{\itbf Y}}_b^\star) \hat{\vec{\itbf
        Y}}_b ] \cdot (\vec\bX_b^\star-\vec\bX_b) }{1-(\hat{\vec{\itbf
      Y}}_b\cdot \hat{\vec{\itbf Y}}_b^\star)^2},
\end{eqnarray*}
so the distance \eqref{def:distance} can be computed explicitly.

\section{Numerical simulations}
\label{sect:num}
In this section we present some simple numerical simulations in order
to illustrate the feasibility of the imaging algorithm \ref{alg:4}.
By simple we mean that the scattered wave field is generated with the
model \eqref{eq:form9}, for spherical particles of radius $a$, using
the Rayleigh-Gans approximation \eqref{eq:form11} of the scattering
kernel.

We consider a laser beam with radius $r_o = 0.5$m, at wavelength $\la
= 1\mu$m, and a Poisson cloud with intensity $\rho= 2.5$m${}^{-3}$, to obtain an
order of $30{,}000$ particles in the beam, up to the range of $1000$m. 

We use up to four groups of cameras, centered at
$\vec{\bX}^{(j)}$, for $j = 1, \ldots, 4$. In the reference system of
coordinates of our computations, with basis denoted by $(\hat{\vec {\itbf r}}_1,\hat{\vec {\itbf
    r}}_2,\hat{\vec {\itbf r}}_3)$, these
locations and the beam axis are
\begin{align*}
\vec{\bX}_b &=(0,0,-1000) , \qquad \hat{\vec{\itbf Y}}_b =(0,0,1),\\
\vec{\bX}^{(1)} &= (100,0,0),\\
\vec{\bX}^{(2)} &= (100\cos(\pi/4),100\sin(\pi/4),-100), \\
\vec{\bX}^{(3)} &= (100 \cos(\pi/3),-100\sin(\pi/3),100) , \\
\vec{\bX}^{(4)} &= (100\cos(\pi/6),100\sin(\pi/6),-300) ,
\end{align*}
with units in meters.

 For simplicity, we assume the same basis
$(\hat{\vec {\itbf e}}_1,\hat{\vec {\itbf e}}_2,\hat{\vec {\itbf
    e}}_3)$ for all four groups of cameras, obtained by the
following rotation of the reference basis
\[
\hat{\vec{{\itbf e}}}_q = \begin{pmatrix} 
1 & 0 & 0 \\
0 & \cos \alpha_1 & -\sin \alpha_1 \\
0 & \sin \alpha_1 & \cos \alpha_1 \end{pmatrix}
 \begin{pmatrix} 
\cos \alpha_2 & 0 & \sin \alpha_2\\0 & 1 & 0 \\
-\sin \alpha_2 &0 &  \cos \alpha_2 \end{pmatrix}
 \begin{pmatrix} 
 \cos \alpha_3& -\sin \alpha_3 & 0 \\
\sin \alpha_3 & \cos \alpha_3 & 0 \\
0 & 0 & 1 \end{pmatrix}
\hat{\vec{\itbf r}}_q,
\]
for $q = 1, 2, 3$, where $\alpha_1 = \pi/7$, $\alpha_2 = \pi/9$, and
$\alpha_3 = \pi/5$.

\begin{figure}[t]
\begin{center}
\hspace{-0.14in} \includegraphics[width=0.27\textwidth]{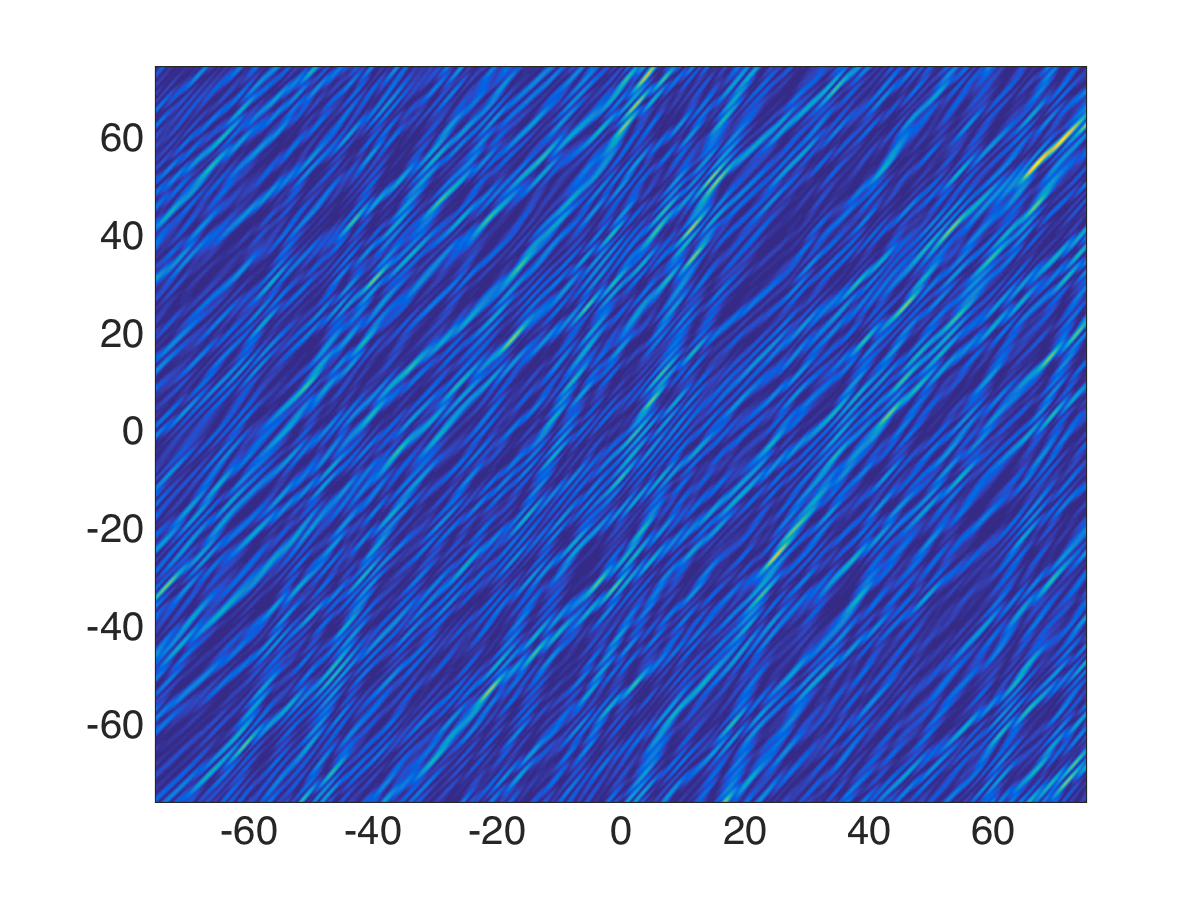}
\hspace{-0.14in}\includegraphics[width=0.27\textwidth]{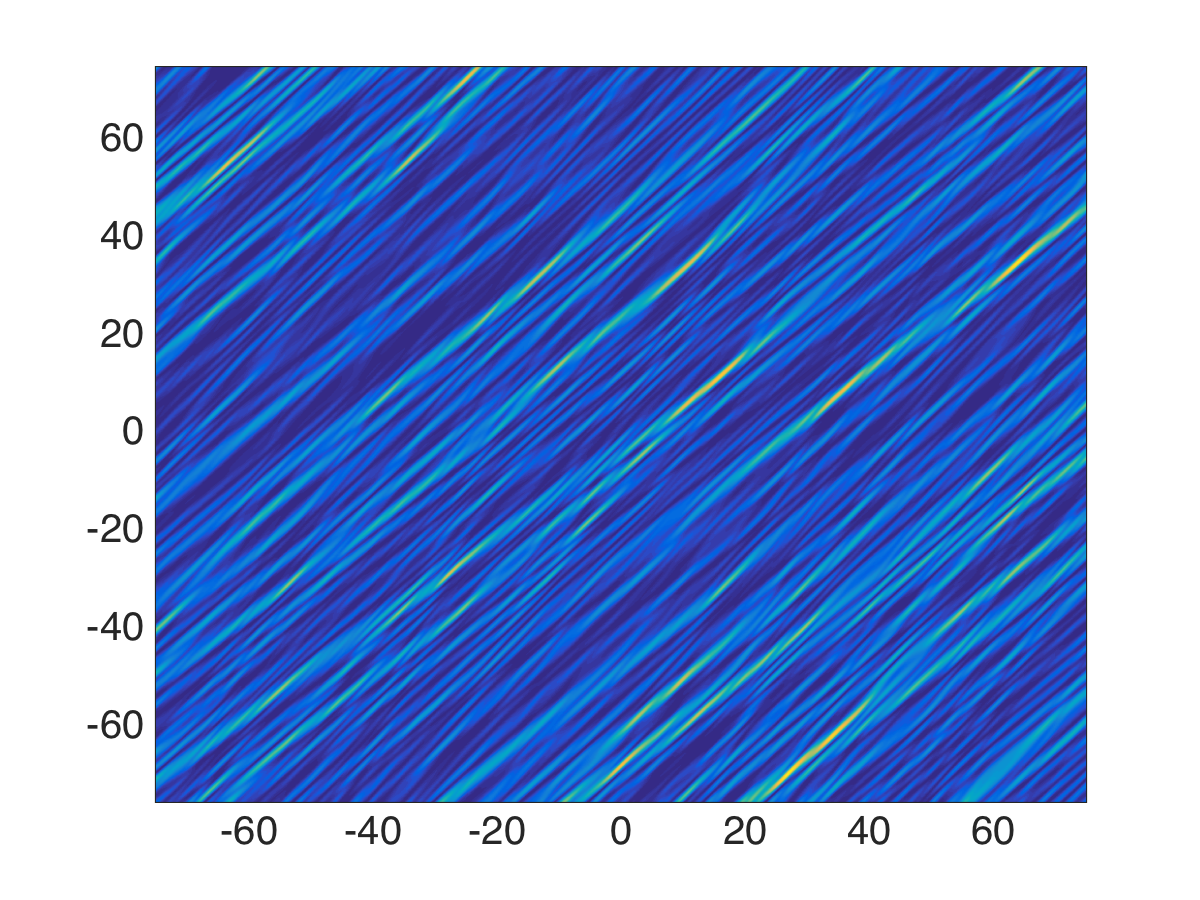}
\hspace{-0.14in}\includegraphics[width=0.27\textwidth]{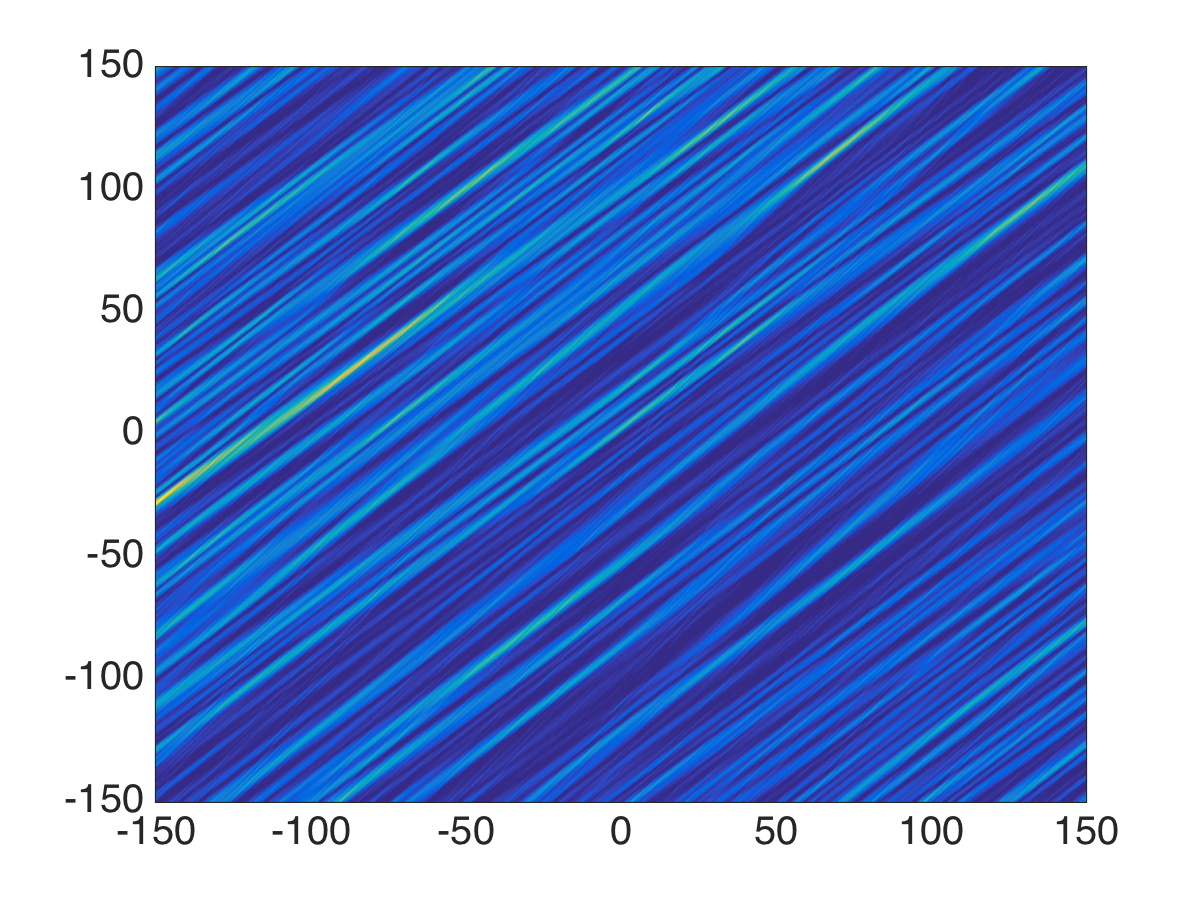}
\hspace{-0.13in}\includegraphics[width=0.27\textwidth]{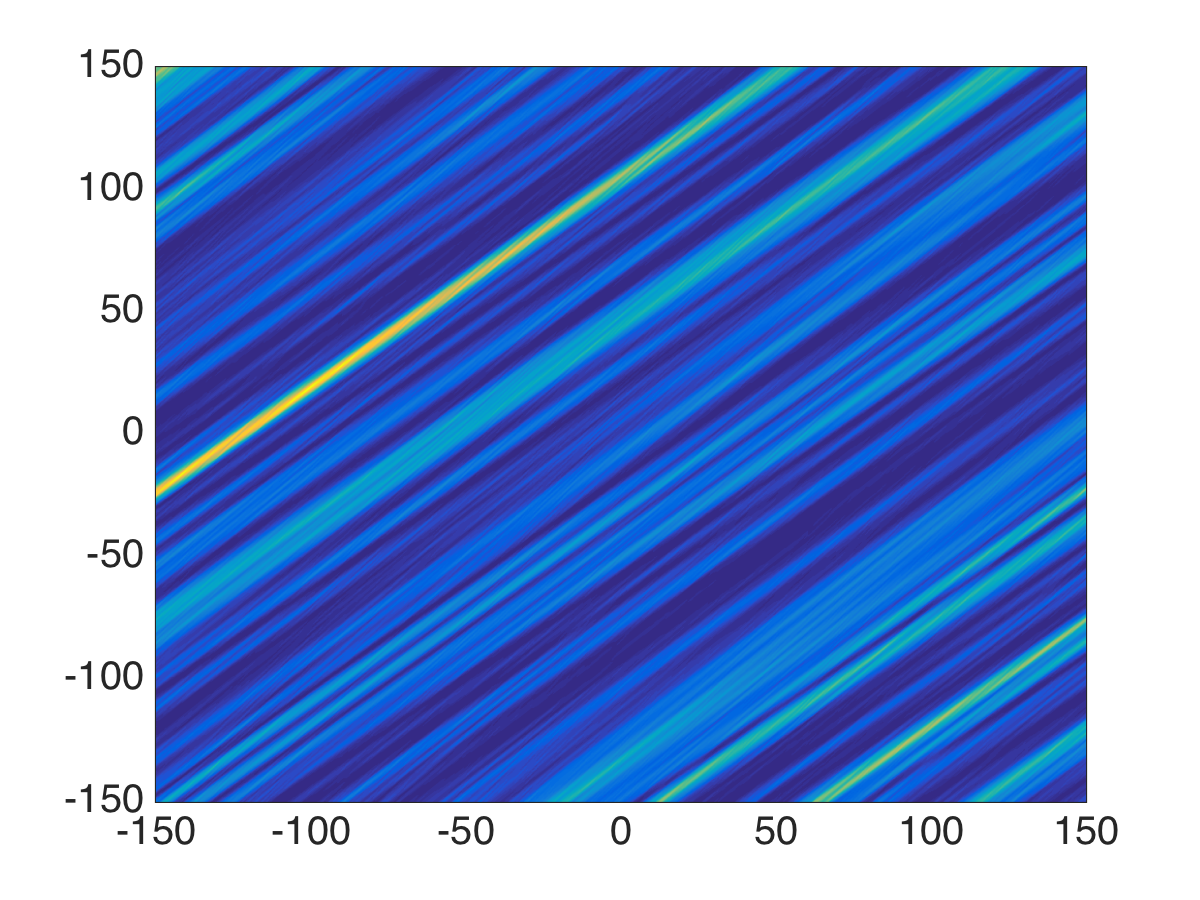}\\
\hspace{-0.14in}\includegraphics[width=0.27\textwidth]{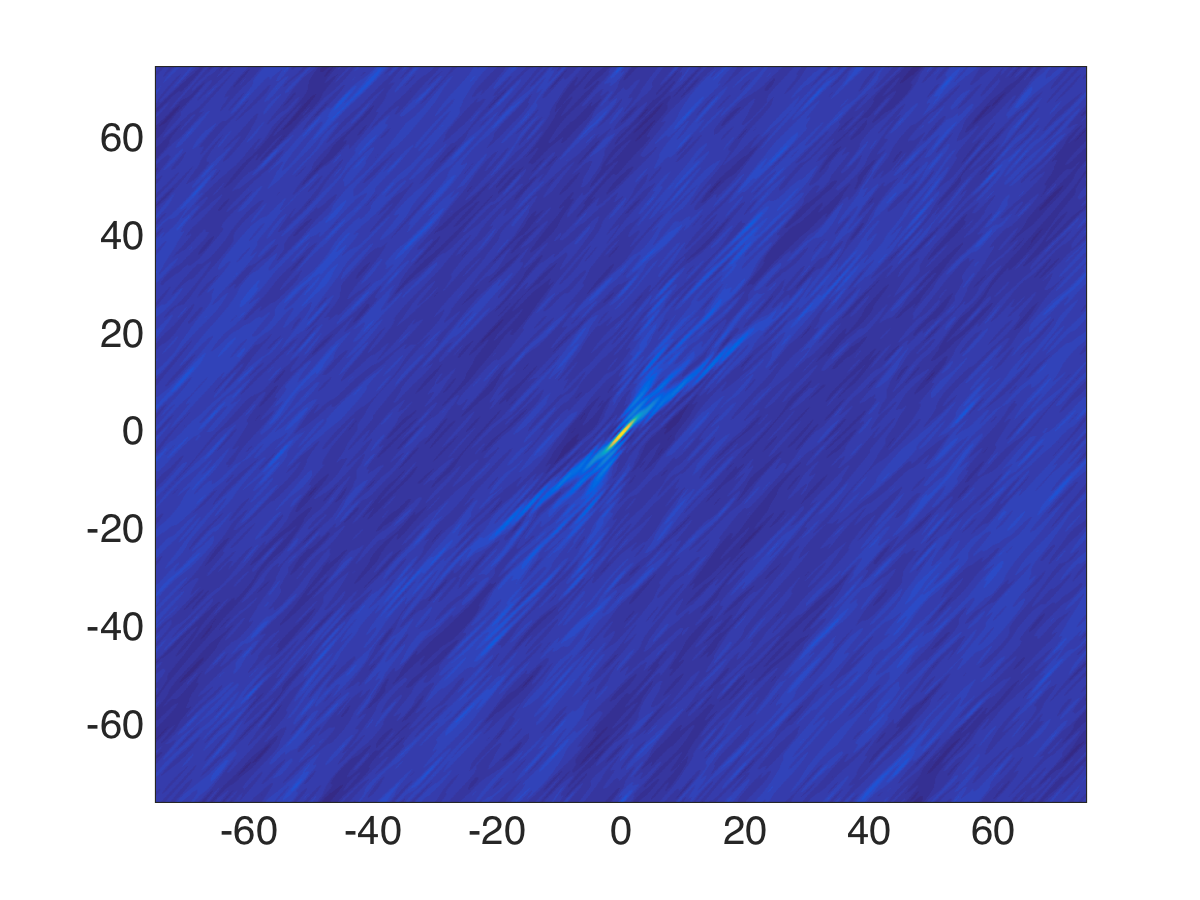}
\hspace{-0.14in}\includegraphics[width=0.27\textwidth]{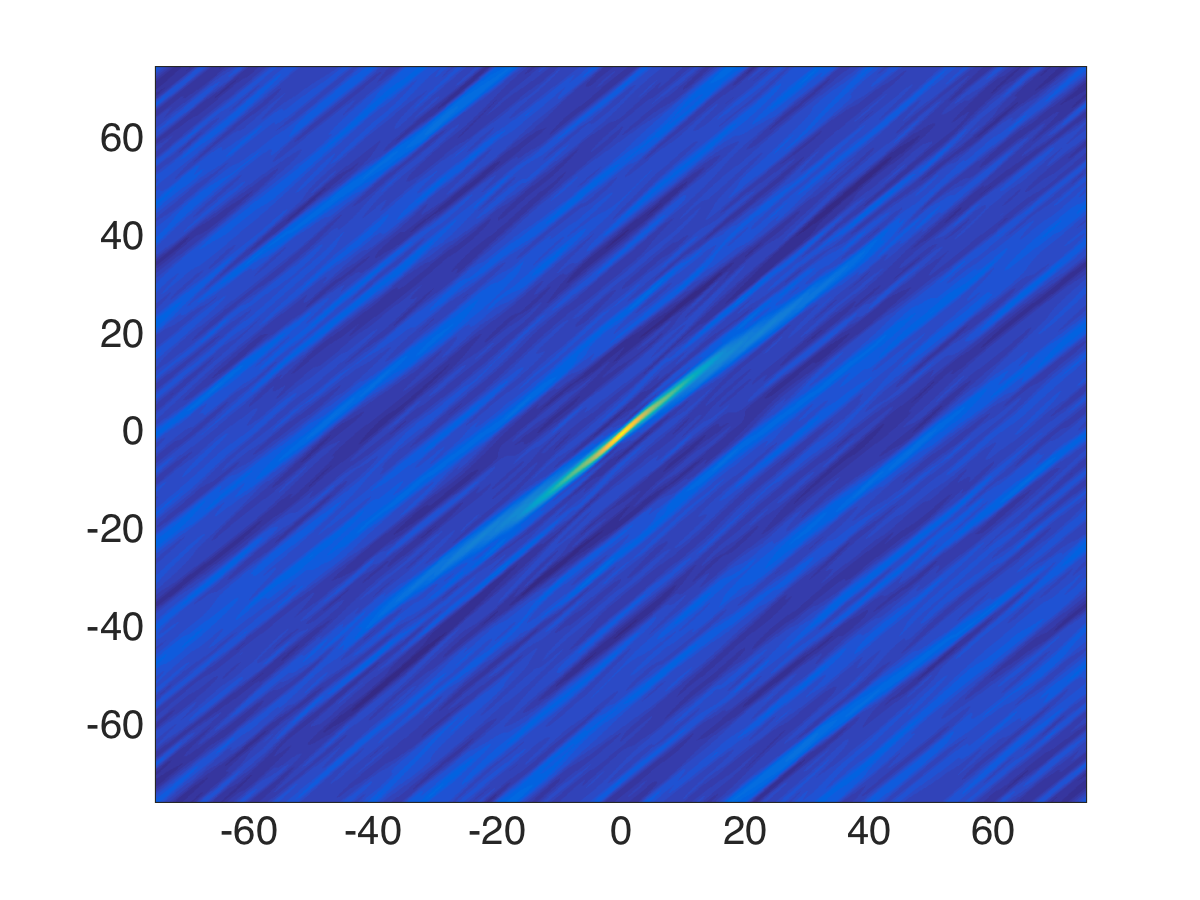}
\hspace{-0.14in}\includegraphics[width=0.27\textwidth]{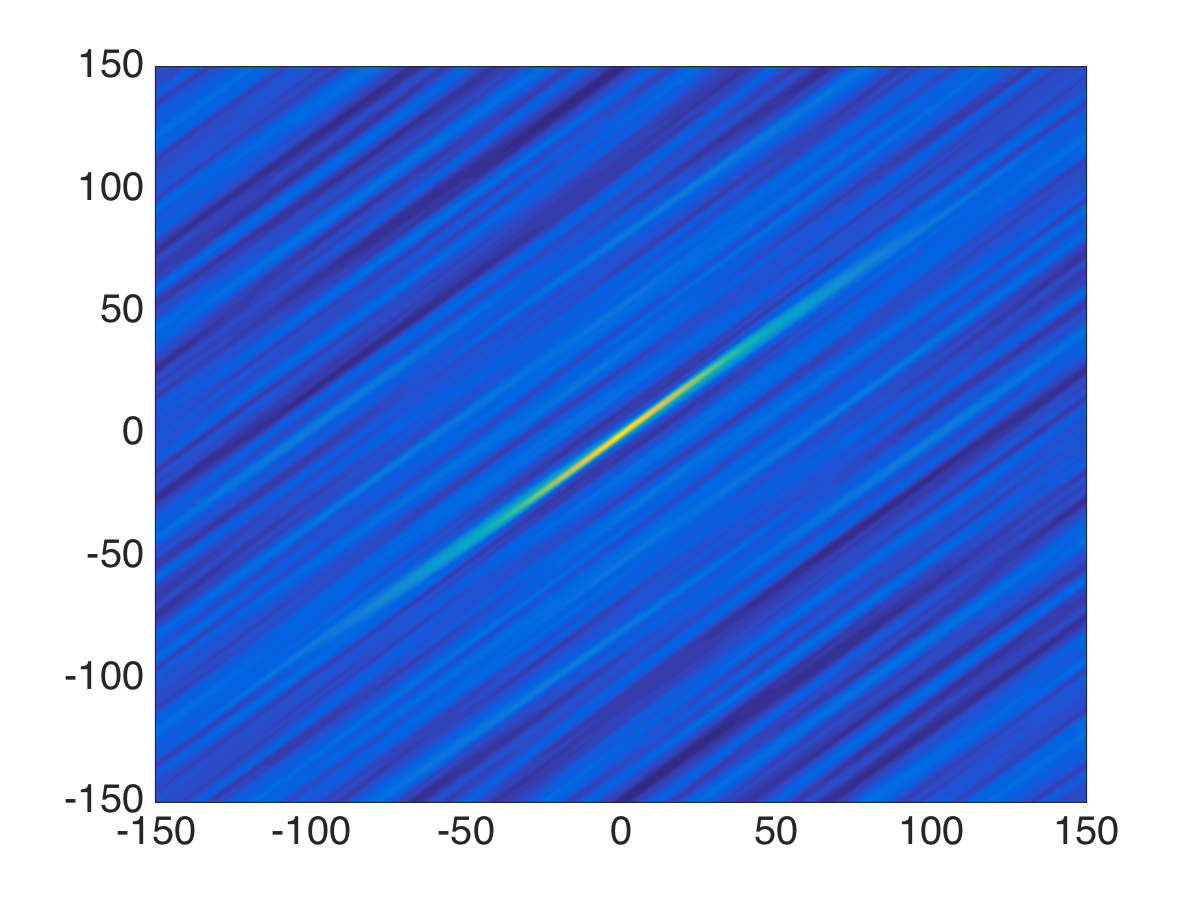}
\hspace{-0.13in}\includegraphics[width=0.27\textwidth]{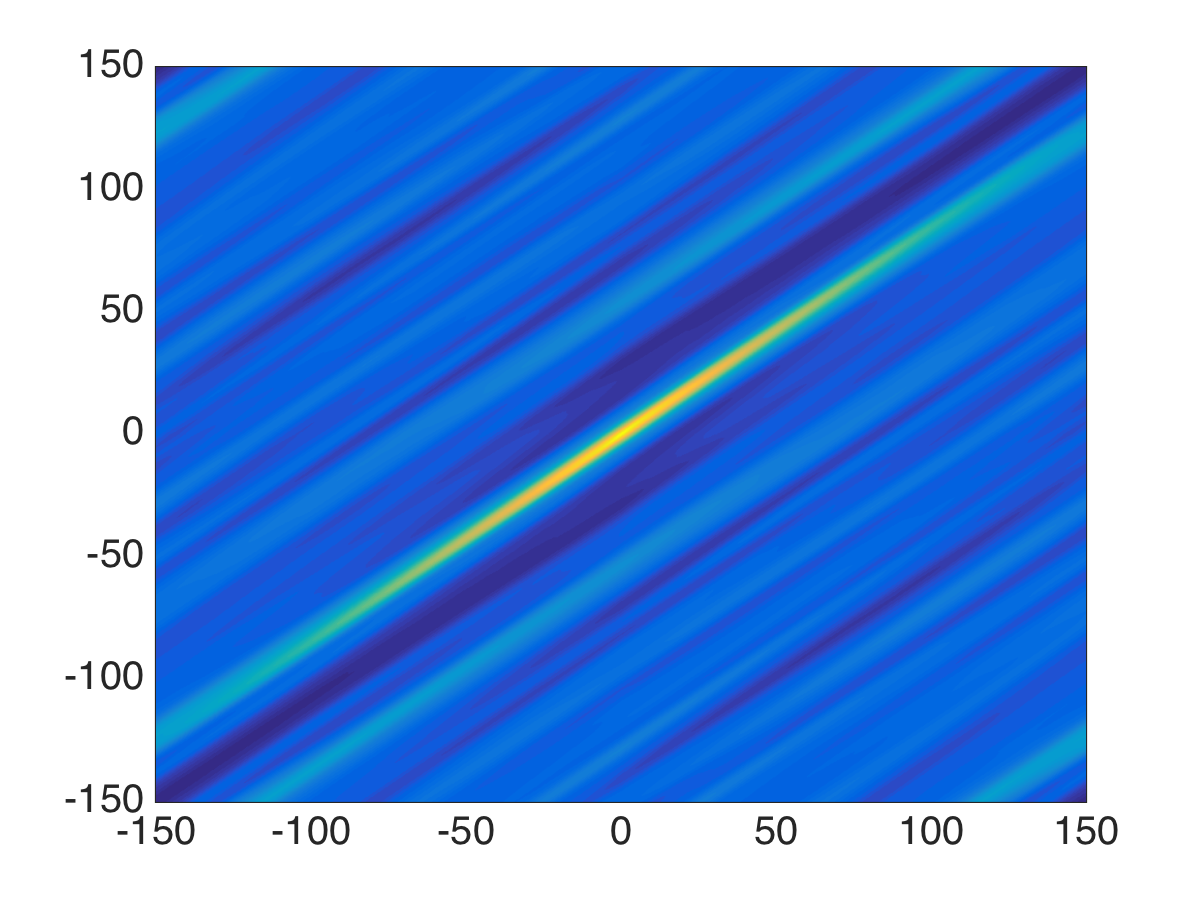}\end{center}
\vspace{-0.1in}
\caption{Top row: The intensity at the first camera centered at
  $\vec{\bX}^{(1)}$, in the plane orthogonal to $\hat{\vec{\itbf
      e}}_3$. Bottom row: The correlation function
  \eqref{eq:CorrelationFunc}. From left to right: $a = 0.1\lambda$, $a
  = 0.5\lambda$, $a = \lambda$, and $a = 2 \lambda$. The axes are pixel
  coordinates in units of the wavelength $\la$.  The colors indicate the values, with yellow the 
  largest  and dark blue the smallest.}
\label{fig:Covar}
\end{figure}
We present results for the following ratios of the radius of the
particle and the wavelength: $a/\la = 0.1, 0.5, 1$, and $2$. In the
cases $a/\la \le 0.5$ we consider an aperture with  diameter $d_A = 150\la$,
and $900$ pixels, to obtain a resolution of $\la/6$. In the other two cases we
have $d_A = 300\la$ and the same number of pixels, to obtain a resolution of
$\la/3$.

We display in Figure \ref{fig:Covar} the speckle pattern of the
intensity at the first camera centered at $\vec\bX^{(1)}$, and the
estimated correlation function \eqref{eq:CorrelationFunc} calculated
as in \eqref{eq:inv4}, using FFT. We note the anisotropy of the decay
of the correlation function, as discussed in section \ref{sect:levelset}.

The matrices ${\boldsymbol{\mathfrak{U}}}^{(j)}$, for $j =
1, \ldots, 4$, are estimated as described in section \ref{sect:onegroup}, for the level set
parameter $\mathfrak{L} = 0.15$. The symmetric  matrix 
${\bf V}$ in the optimization \eqref{eq:optimize} is represented by three search parameters: the two entries on the diagonal and the one off the diagonal. The minimization is solved with the MATLAB routine \emph{fmincon}. 

We display in the next table the eigenvalues $(\Lambda_q^{(1)})_{1\le q \le 3}$ of
${\boldsymbol{\mathfrak{U}}}^{(1)}$, for the four ratios $a/\la$. The eigenvalues of the 
other matrices have a similar behavior.

\vspace{0.08in}
\begin{center}
  \begin{tabular}{|l|c|c|r|}
    \hline
    $a/\la$ & $\Lambda_1^{(1)}$ & $\Lambda_2^{(1)}$ & $\Lambda_3^{(1)}$ \\ \hline 
    0.1 & 29.49 & 1.14 & -0.10 \\ \hline 
    0.5 & 11.86 & 0.26 & 0.02 \\ \hline
    1 & 2.69 & 0.01 & -0.05 \\ \hline
    2 & 0.5417 & 0.004 & -0.02 \\\hline
  \end{tabular}
\end{center}

\vspace{0.08in}
\noindent Recall that $\Big(\Lambda_j^{(1)}\Big)^{-1/2}$, $j=1,2,3$, determine the
semi-principal axes of the ellipsoid which approximates the level set
of the correlation function, at level value $1 -\mathfrak{L}$. These
are proportional to the correlation lengths, and as stated in
Propositions \ref{prop:small} and \ref{prop:large}, the correlation
lengths increase with the size of the particles. This is why
$\Lambda_1^{(1)}$ is the largest when $a = 0.1\la$.

The smallest eigenvalue $\Lambda_3^{(1)}$ is well separated from the others
in the case $a/\la = 0.1$, so we use Algorithm \ref{alg:1}
to estimate the laser beam in this case. In all other cases the
estimation is done using Algorithm \ref{alg:2}  for two groups of 
cameras, and Algorithm \ref{alg:4} for $3$ or $4$ groups of
cameras. We do not use Algorithm \ref{alg:1} for the case $a = 0.5
\la$, because the gap $\Lambda_2^{(2)}-\Lambda_1^{(1)}$ was smaller for the matrix
${\boldsymbol{\mathfrak{U}}}^{(2)}$. 

The results obtained with the first two groups of cameras are in the next table, where we recall the
estimation error quantifiers $\beta$ and $d$ are defined in \eqref{def:angle} and
\eqref{def:distance}. The angle $\beta$ is in degrees and the distance
$d$ is in meters.

\vspace{0.08in}
\begin{center}
  \begin{tabular}{|l|c|c|}
    \hline
    $a/\la$ & $\beta$ & $d$ \\ \hline 
    0.1 & $3.5^o$ & $2.47$m\\ \hline 
    0.5 & $30.5^o$ & $9.27$m \\ \hline
    1 & $17.6^o$ & $6.12$m \\ \hline
    2 & $12.6^o$ & $3.94$m\\\hline
  \end{tabular}
\end{center}

\vspace{0.08in} \noindent We note that the best estimate is for the smallest
particles.  This is expected from scenario~1 
discussed in section \ref{sect:twogroup}, because the eigenvectors $\hu_3^{(j)}$ are robust to estimation errors, 
and approximate well  the unit vectors normal
to the planes defined by the centers of the cameras and the axis of
the laser.  The estimates for the larger particles are worse, with the case
$a = 2 \la$ being the better one, as it is marginally in scenario 2
discussed in section \ref{sect:twogroup}, where the leading
eigenvectors of $\hu_1^{(j)}$, for $j = 1,2$, are
nearly orthogonal to the axis of the laser.

The results improve significantly when we use the three first groups
of cameras, as shown in the next table. We only display the results
for the larger particles, because Algorithm \ref{alg:4} terminates at
Step 1 in the case $a = 0.1\la$.

\vspace{0.08in}
\begin{center}
  \begin{tabular}{|l|c|c|}
    \hline
    $a/\la$ & $\beta$ & $d$ \\ \hline 
    0.5 & $2.6^o$ & $1.84$m \\ \hline
    1 & $2.9^o$ & $2.42$m \\ \hline
    2 & $2.4^o$ & $1.72$m\\\hline
  \end{tabular}
\end{center}

\vspace{0.08in}
\noindent The results with all four groups of cameras are qualitatively the same,
as shown below.
\vspace{0.08in}
\begin{center}
  \begin{tabular}{|l|c|c|}
    \hline
    $a/\la$ & $\beta$ & $d$ \\ \hline 
    0.5 & $2.4^o$ & $1.02$m \\ \hline
    1 & $0.9^o$ & $0.57$m \\ \hline
    2 & $0.9^o$ & $2.08$m\\\hline
  \end{tabular}
\end{center}

\vspace{0.08in}
\section{Summary}
\label{sect:sum}
This paper introduces a novel algorithm for imaging a laser beam using
measurements at CCD cameras that do not lie in the footprint of the
beam. Motivated by the application of detection and
characterization of high energy, continuous wave  lasers in maritime atmospheres, we
consider the light scattered away from the axis of the
beam by small particles suspended in air (aerosols). We derive a model 
of the light intensity at the cameras using the Mie scattering theory for  a Poisson cloud of spherical particles.
This model displays the generic dependence of the   speckle
pattern of the  intensity on the laser beam.  By generic we mean that the 
conclusions extend to mixtures of particle sizes and shapes and to strong scattering regimes, 
as long as the cameras are not farther than a transport mean free path from the 
axis of the laser.   The imaging algorithm is based on the behavior of the correlation 
function of the  intensity, in particular on its anisotropic decay 
on length scales, called correlation lengths, which depend on the orientation of the axis 
of the laser. It estimates this correlation function using measurements 
at groups of three CCD cameras with a common center and apertures that lie in  different planes.
Two such groups of cameras are sufficient for imaging the laser beam 
in the case of either small or large particles with respect to the wavelength. For particles of general 
size, at least three groups of cameras are needed to obtain an accurate image.  The theoretical results
are validated with numerical simulations.

\section*{Acknowledgments}
\label{sec:acknowledge}
This material is based upon research supported in part by the
U.S. Office of Naval Research under award number N00014-17-1-2057 and 
by AFOSR under award number FA9550-15-1-0118.

\appendix
\section{The covariance function of the scattered field
  for small particles}
\label{ap:propsmall}
We begin the proof of Proposition \ref{prop:small} with the model
\eqref{eq:form9} of the scattered field and the approximation
\eqref{eq:form10} of the Mie scattering kernel by the constant $\eta$
defined in \eqref{eq:form10},
\begin{equation}
  u_{\rm s}(\vec\bx) \approx {k^2 \eta} \sum_j G(\vec \bx,\vec\bx_j)
  u_{\rm b}(\vec \bx_j).
\end{equation}
Using the expressions \eqref{eq:form5} and \eqref{eq:form7} of the
laser beam field $u_{\rm b}$ and the Green's function $G$, and then taking
expectations as described in \eqref{eq:expectation}, we obtain
\begin{align}
\hspace{-0.08in}\EE \big[ &u_{\rm s}(\vec\bx) ] \approx \rho k^2 \eta
\int_0^\infty dz' \int_{\RR^2} d \bx' \, {G}(\vec\bx,\vec\bx')
u_{\rm b}(\vec\bx') \nonumber \\ &= \frac{\rho k^2 \eta}{4 \pi}
\int_0^\infty dz' \int_{\RR^2} d \bx'\,
\frac{r_o^2}{|\vec\bx-\vec\bx'| R_{z'}^2} \exp\Big[ -
  \frac{|\bx'|^2}{R_{z'}^2} +(ik-k_{\rm d})( z'+
  |\vec\bx-\vec\bx'|)\Big] . \label{eq:usc}
\end{align}
Here we decomposed $\vec\bx' = (\bx',z')$, with range coordinate $z' >
0$ along the axis of the beam and two-dimensional cross-range vector
$\bx'$.
In our scaling regime
$
k  z' \sim {L_z}/{\la} \gg 1,
$
and the mean zero result
\eqref{eq:prop1} follows from the Riemann-Lebesgue lemma, due to the
rapid phase $\exp(ikz')$ in equation \eqref{eq:usc}.

Next we let $\vec\bx_1=\vec\bX + \vec\bx/2$ and $\vec\bx_2=\vec\bX -
\vec\bx/2,$ and calculate the covariance function 
\begin{align}
\nonumber \EE \big[ &u_{\rm s}(\vec\bx_1)\overline{u_{\rm s}}(\vec\bx_2) \big]
= \eta^2 k^4 \rho \int_0^\infty d z'\int_{\RR^2} d
\bx'\, {G}(\vec\bx_1,\vec\bx') \overline{{G}(\vec\bx_2,\vec\bx')}
|u_{\rm b}(\vec\bx')|^2 \\ \nonumber &= \frac{ \eta^2 k^4 \rho
}{16\pi^2} \int_0^\infty dz' \int_{\RR^2}d \bx'\,
\frac{r_o^4}{|\vec\bx_1-\vec\bx'||\vec\bx_2-\vec
  \bx'||R_{z'}|^4}\exp\Big( - \frac{2 r_o^2 |\bx'|^2}{|R_{z'}|^4} -
2k_{\rm d} z' \Big) \\ &
\quad \times \exp\Big[ i k \big(
  |\vec\bx_1-\vec\bx'| - |\vec\bx_2-\vec\bx'| \big) -k_{\rm d} \big(
  |\vec\bx_1-\vec \bx'| + |\vec\bx_2-\vec\bx'| \big)\Big] ,
\label{eq:A2}
\end{align}
where we used that
\[{\rm Re}\left( \frac{1}{R_{z'}^2}\right) = \frac{r_o^2}{|R_{z'}|^4}.\]
We have
\begin{align}
  k \big( |\vec\bx_1-\vec\bx'| - |\vec\bx_2-\vec\bx'| \big) &= k \Big(
  \Big|\vec\bX-\vec\bx' + \frac{\vec\bx}{2}\Big| -
  \Big|\vec\bX-\vec\bx' - \frac{\vec\bx}{2}\Big| \Big)\nonumber \\ &=
  k \vec \bx \cdot \frac{(\vec \bX - \vec \bx')}{|\vec \bX - \vec
    \bx'|} + O\Big(\frac{k |\vec\bx|^3}{|\vec \bX - \vec
    \bx'|^2}\Big)
%     \nonumber \\ &\approx k \vec \bx \cdot \frac{(\vec
%    \bX - \vec \bx')}{|\vec \bX - \vec \bx'|}
, \label{eq:A3}
\end{align}
where the residual is negligible if 
\begin{align}
\label{eq:A3b}
|\vec\bx|  \ll \lambda^{1/3} L_\bx^{2/3},
\end{align}
which we assume.
%\begin{align*}
%  \frac{k |\vec\bx|^3}{|\vec \bX - \vec
%    \bx'|^2} \sim \frac{A^3}{\la L_\bx^2} \ll 1.
%\end{align*}
We also have the approximations for the amplitude factors
\begin{equation}
  \frac{1}{|\vec\bx_1-\vec\bx'|} \approx
  \frac{1}{|\vec\bx_2-\vec\bx'|} \approx \frac{1}{|\vec \bX - \vec
    \bx'|} \approx \frac{1}{\sqrt{|\bX|^2 + (Z-z')^2}},
     \label{eq:A3a}
\end{equation}
and 
\begin{equation}
  k_{\rm d} \big( 2 z' + |\vec\bx_1-\vec\bx'| + |\vec\bx_2-\vec\bx'|
  \big) \approx 2 k_{\rm d} [z' + \sqrt{|\bX|^2 + (Z-z')^2}] \approx 2
  k_{\rm d} (z'+ |Z-z'|),
\label{eq:A3p}
\end{equation}
by assumption \eqref{eq:form13} and the scaling relations 
\[
|\bx'| \sim r_o \ll |\bX| \sim L_\bx \ll Z \sim L_z.
\]

Substituting \eqref{eq:A3}, \eqref{eq:A3a}, and \eqref{eq:A3p} into equation
\eqref{eq:A2} we obtain
\begin{align}
\nonumber \EE \big[ u_{\rm s}(\vec\bx_1)\overline{u_{\rm s}}(\vec\bx_2) \big]
\approx & \frac{ \eta^2 k^4 \rho }{16\pi^2} \int_0^{\infty} \hspace{-0.05in} dz' e^{-2k_{\rm d}(z'+|Z-z'|)}\int_{\RR^2} d
\bx' \frac{r_o^4 }{\big[|\bX|^2 +(Z-z')^2\big]|R_{z'}|^4} \\ &\times
\exp\Big[- \frac{2r_o^2 |\bx'|^2}{|R_{z'}|^4}+ i k\frac{(\bX -\bx')
    \cdot \bx +(Z-z') z}{\sqrt{|\bX-\bx'|^2 +(Z-z')^2}}
  \Big] \label{eq:A4}.
\end{align}

The integrand in this equation is large for $|\bx'| \lesssim r_o$ and we have 
%$|\vec\bx|\lesssim {\rm diam}(A)$  and 
$|\bX|\sim L_\bx$, so
we can expand the phase as
\begin{align}
  k\frac{(\bX -\bx') \cdot \bx +(Z-z') z}{\sqrt{|\bX-\bx'|^2
      +(Z-z')^2}} =& k\frac{\bX  \cdot \bx +(Z-z')
    z}{\sqrt{|\bX|^2 +(Z-z')^2}} \nonumber \\
 \nonumber
    &+ k \bx' \cdot  
    \frac{[ \bX\cdot \bx +(Z-z')z] \bX - [|\bX|^2+(Z-z')^2] \bx }{[|\bX|^2+(Z-z')^2]^{3/2}}
    \\
    &+
    O\Big( k\frac{|\bx'|^2|\vec\bx|}{|\bX|^2}\Big) . \label{eq:phase}
\end{align}
Note that the residual is small if 
\begin{align}
\label{eq:A3bb}
|\vec\bx| \ll \lambda (L_\bx^2 / r_o^2) ,
\end{align}
which we assume.
This allows us to integrate over $\bx'$ in (\ref{eq:A4}) and we  obtain
\begin{align}
\nonumber \EE \big[ u_{\rm s}(\vec\bx_1)\overline{u_{\rm s}}(\vec\bx_2) \big]
\approx& \frac{ \eta^2 k^4 \rho r_o^2 }{32\pi} \int_0^\infty dz'
\frac{e^{-2k_{\rm d}(z'+|Z-z'|)}}{\big[|\bX|^2 +(Z-z')^2\big]}
\\ &\times \exp\Big[ i k\frac{\bX \cdot \bx +(Z-z') z}{\sqrt{|\bX|^2
      +(Z-z')^2}}  - \frac{k^2 |R_{z'}|^4 {\cal Q}(\vec\bx, \vec\bX,z')}{8 r_o^2} \Big] dz' , \label{eq:A5}
\end{align}
with
\begin{align*}
{\cal Q}(\vec\bx, \vec\bX,z') =& 
\Big|
 \frac{[ \bX\cdot \bx +(Z-z')z] \bX - [|\bX|^2+(Z-z')^2] \bx }{[|\bX|^2+(Z-z')^2]^{3/2}}
\Big|^2  \\
=
&
\frac{-(\bx\cdot\bX)^2 |\bX|^2 +z^2 (Z-z')^2 |\bX|^2 
+|\bx|^2 [|\bX|^2+(Z-z')^2]^2}{[|\bX|^2+(Z-z')^2]^3}\\
&- \frac{2(\bx\cdot\bX) [ \bx\cdot\bX +(Z-z')z] (Z-z')^2}{[|\bX|^2+(Z-z')^2]^3}  .
\end{align*}
Now let us change variables
\begin{equation}
  \label{eq:changez}
z' = Z - |\bX| \zeta, \quad \zeta \in \Big(-\infty,
\frac{Z}{|\bX|}\Big),
\end{equation}
and use that $Z/|\bX| \gg 1$  to extend
the $\zeta$ interval to the real line. Equation \eqref{eq:A5} becomes
\begin{align}
\nonumber \EE \big[ u_{\rm s}(\vec\bx_1)\overline{u_{\rm s}}(\vec\bx_2) \big]
\approx & \frac{ \eta^2 k^4 \rho r_o^2 e^{-2k_{\rm d} Z}}{32\pi |\bX|}
\int_{-\infty}^\infty \frac{e^{-2 k_{\rm d} |\bX| (|\zeta| -
    \zeta)}}{1 +\zeta^2} \\ &\times \exp\Big(i k\frac{\hat{\bX} \cdot
  \bx + \zeta z}{\sqrt{1 +\zeta^2}} - \frac{k^2 |R_{_{Z - |\bX|\zeta}}|^4
 \tilde{\cal Q}(\vec\bx,\hat\bX,\zeta) }{8 |\bX|^2 r_o^2} \Big) d\zeta , \label{eq:A6}
\end{align}
where $\hat{\bX} = \bX/|\bX|$ and
$$
\tilde{\cal Q}(\vec\bx, \hat\bX,\zeta) = \frac{-(\bx\cdot\hat\bX)^2   (1+2\zeta^2) 
+|\bx|^2  (1+\zeta^2)^2 + z^2  \zeta^2  -2 (\bx\cdot\hat\bX) z   \zeta^2}{
 (1+\zeta^2)^3}  .
$$
 We can simplify this result further, by
noting that since $1/(1+\zeta^2)$ is integrable, only the terms with $|\zeta| = O(1)$
contribute to value of the integral (\ref{eq:A6}), and we can approximate
\[
|R_{_{Z -
    |\bX|\zeta}}| \approx r_o,
\qquad 
e^{-2 k_{\rm d} |\bX| (|\zeta| - \zeta)} \approx 1, 
\]
by the scaling assumptions \eqref{eq:form14} and \eqref{eq:form13}.
Furthermore, $|\hat{\bX} \cdot \bx| $ and $|z|$ should be at most of order 
$\lambda$ otherwise the integral in (\ref{eq:A6}) averages out by the Riemann-Lebesgue lemma,
and in these conditions
$$
\frac{k^2 r_o^2 }{8  |\bX|^2}  \tilde{\cal Q}(\vec\bx, \hat\bX,\zeta) = \frac{k^2 r_o^2 }{8  |\bX|^2}  | \hat{\bX}^\perp \cdot \bx|^2
+o(1)
,
$$
using the orthonormal basis $\{\hat \bX, \hat{\bX}^\perp\}$ in the
cross-range plane,
so that
\begin{align}
\nonumber \EE \big[ u_{\rm s}(\vec\bx_1)\overline{u_{\rm s}}(\vec\bx_2) \big]
\approx & \frac{ \eta^2 k^4 \rho r_o^2 e^{-2k_{\rm d} Z}}{32\pi |\bX|}
\int_{-\infty}^\infty \frac{1}{1 +\zeta^2} \\ &\times \exp\Big(i k\frac{\hat{\bX} \cdot
  \bx + \zeta z}{\sqrt{1 +\zeta^2}} - \frac{k^2 r_o^2 | \hat{\bX}^\perp \cdot \bx|^2
 }{8  |\bX|^2 (1+\zeta^2)} \Big) d\zeta  .
  \label{eq:A6b}
\end{align}
Equations~(\ref{eq:cov1}-\ref{eq:cov1p}) follow after one more
change of variables,
\begin{equation}
  \zeta/\sqrt{1+\zeta^2}=\cos \alpha.
\end{equation}
Note that this result shows that the covariance function, as a function of $\vec\bx$,
has the form
of an anisotropic peak centered at ${\bf 0}$ with radii of the order of $\lambda$ in the $z$- and
$\hat{\bX}$-direction, and of the order of $\lambda |\bX|/r_o$ in the $\hat{\bX}^\perp$-direction.
In order to see this peak, the radius of the camera should be ideally larger than $\lambda |\bX|/r_o$.
If it is not as large, then the elongated peak will extend to the boundary of the domain.

Note also that (\ref{eq:A3b}) and (\ref{eq:A3bb}) are readily fulfilled for $|\vec\bx|\lesssim \lambda$.
When the diameter $d_A$ is larger than $\lambda |\bX|/r_o$, 
these conditions are fulfilled for all $\vec\bx$ in the peak of the 
covariance function if they hold for all $|\vec\bx|\leq \lambda L_\bx/r_o$, which happens
if $(\lambda L_\bx/r_o)^3 \ll \lambda L_\bx^2$ 
and $\lambda L_\bx/r_o \ll \lambda(L_\bx^2 /r_o^2) $.
This imposes the additional condition $   \lambda^2 L_\bx \ll r_o^3$.

\section{Proof of the Gaussian summation rule}
\label{ap:Gaussian}
We prove here that in our scaling regime the fourth order moments of
the scattered field satisfy approximately the Gaussian summation rule.
This gives the result stated in Proposition \ref{prop:Gauss}.

Let $\vec\bx_1=\vec\bX + \vec\bx/2$ and $\vec\bx_2=\vec\bX -
\vec\bx/2$ be two points in the camera, and use equation 
(\ref{eq:camp2b}) to write  the second moment of the intensity
\begin{align}
\nonumber \EE \big[ |u_{\rm s}(\vec\bx_1)|^2 |{u_{\rm s}}(\vec\bx_2) |^2\big]
&= \EE \big[ |u_{\rm s}(\vec\bx_1)|^2\big] \EE \big[ |{u_{\rm s}}(\vec\bx_2)
  |^2\big] + \big|\EE \big[
  u_{\rm s}(\vec\bx_1)\overline{u_{\rm s}}(\vec\bx_2) \big] \big|^2\nonumber
\\ &+ \big| \EE \big[ u_{\rm s}(\vec\bx_1) {u_{\rm s}}(\vec\bx_2) \big]
\big|^2 + \mathcal{R}. \label{eq:4Mom}
\end{align}
Here $\mathcal{R}$ is the residual given by the last term in
\eqref{eq:camp2b},
\begin{align*}
\mathcal{R} &= \rho \eta^4 k^8 \int_0^\infty dz' \int_{\RR^2} d \bx'
|{G}(\vec\bx_1,\vec\bx')|^2|{G}(\vec\bx_2,\vec\bx')|^2
|u_{\rm b}(\vec\bx')|^4 \\ &\approx \frac{\rho \eta^4 k^8 }{(4\pi)^4}
\int_0^\infty \hspace{-0.05in} dz' \int_{\RR^2} d \bx' \frac{
e^{-4 k_{\rm d} (z'+|Z-z'|)}}{ [
    |\bX|^2 +(Z-z')^2]^2} \frac{r_o^8}{|R_{z'}|^8} \exp\Big(- \frac{4
  r_o^2 |\bx'|^2}{|R_{z'}|^4} \Big) \\ &= \frac{\rho \eta^4 k^8 r_o^6
  }{4^5 \pi^3} \int_0^\infty dz' \frac{e^{-4 k_{\rm d} (z'+|Z-z'|)}}{ [ |\bX|^2
    +(Z-z')^2]^2 |R_{z'}|^4},
\end{align*}
where we used the same approximation as in \eqref{eq:A3p}.
With the change of variables \eqref{eq:changez}, and using that
$Z/|\bX| \gg 1$, we estimate the residual by
\begin{align*}
  \mathcal{R} \approx \frac{\rho \eta^4 k^8 r_o^2 e^{-4 k_{\rm d}
      Z}}{4^5 \pi^3 |\bX|^3} \int_{-\infty}^\infty \frac{1}{ (1
    +\zeta^2 )^2} d\zeta = \frac{\rho \eta^4 k^8 r_o^2 e^{-4 k_{\rm d}
      Z}}{2^{11} \pi^2 |\bX|^3}. \label{eq:ResEst}
\end{align*}

Let us compare ${\mathcal R}$ with the first two terms in
\eqref{eq:4Mom}, which are of the same order, estimated from equation
\eqref{eq:meanInten},
\begin{align*}
\big|\EE \big[ u_{\rm s}(\vec\bx_1)\overline{u_{\rm s}}(\vec\bx_2) \big]
\big|^2 \sim \EE \big[ |u_{\rm s}(\vec\bx_1)|^2\big] \EE \big[
  |{u_{\rm s}}(\vec\bx_2) |^2\big] \approx \left(\frac{ \eta^2 k^4 \rho
  r_o^2 e^{-2k_{\rm d} Z}}{32 |\bX|}\right)^2.
\end{align*}
We obtain that 
\begin{align}
  \frac{\mathcal{R}}{\EE \big[ |u_{\rm s}(\vec\bx_1)|^2\big] \EE \big[
      |{u_{\rm s}}(\vec\bx_2) |^2\big]} \approx \frac{1}{2 \pi^2 \rho
    r_o^2 |\bX|} \ll 1,
\end{align}
because in our scaling
\begin{equation}
  \rho r_o^2 |\bX| = \frac{r_o^2 |\bX|}{\ell^3} \gg 1.
\end{equation}
Therefore, the residual is negligible in \eqref{eq:4Mom} and the
covariance is given by
\begin{equation}
{\rm Cov} \big( |u_{\rm s}(\vec\bx_1)|^2 , |{u_{\rm s}}(\vec\bx_2)
  |^2\big) \approx \big|\EE \big[
  u_{\rm s}(\vec\bx_1)\overline{u_{\rm s}}(\vec\bx_2) \big] \big|^2  +
\big|\EE \big[
  u_{\rm s}(\vec\bx_1){u_{\rm s}}(\vec\bx_2) \big] \big|^2.
\label{eq:B4} 
\end{equation}

It remains to show that the last term in \eqref{eq:B4} is small. We
have from the model \eqref{eq:form9} and equation
\eqref{eq:expectation} that 
  \begin{align}
\nonumber \EE \big[ &u_{\rm s}(\vec\bx_1){u_{\rm s}}(\vec\bx_2) \big]
= \eta^2 k^4 \rho \int_0^{\infty}\hspace{-0.05in}d z'\int_{\RR^2} d
\bx'\, {G}(\vec\bx_1,\vec\bx') {{G}(\vec\bx_2,\vec\bx')}
|u_{\rm b}(\vec\bx')|^2 \\ \nonumber &= \frac{ \eta^2 k^4 \rho
}{16\pi^2} \int_0^{Z_{\rm P}} \hspace{-0.05in}dz' \int_{\RR^2}d \bx'\,
\frac{r_o^4}{|\vec\bx_1-\vec\bx'||\vec\bx_2-\vec
  \bx'||R_{z'}|^4}\exp\Big[ - \frac{2 r_o^2 |\bx'|^2}{|R_{z'}|^4} +
2(i k-k_{\rm d}) z' \Big] \\ &\times \exp\Big[ i k \big(
  |\vec\bx_1-\vec\bx'| + |\vec\bx_2-\vec\bx'| \big) -k_{\rm d} \big(
  |\vec\bx_1-\vec \bx'| + |\vec\bx_2-\vec\bx'| \big)\Big].
\label{eq:B5}
\end{align}
This expression contains very large phases: $2 k z' \gg 1$ and $k
\big( |\vec\bx_1-\vec\bx'| + |\vec\bx_2-\vec\bx'| \big) \gg 1$,
so the result is small after integration, by the Riemann-Lebesgue
lemma. The result stated in Proposition \ref{prop:Gauss} follows.

\end{document}